\documentclass[
showpacs,preprintnumbers,amsmath,amssymb,nofootinbib]{revtex4}

\usepackage{amsmath}
\usepackage{graphicx}
\usepackage{amsfonts}
\usepackage{array}
\usepackage{amsthm}

\usepackage{latexsym}

\usepackage{epic}
\usepackage{eepic}
\usepackage{graphicx}

\begin{document}


\title{Extremal dyonic black holes in D=4 Gauss-Bonnet gravity}

\author{Chiang-Mei Chen} \email{cmchen@phy.ncu.edu.tw}
\affiliation{Department of Physics and Center for Mathematics and
Theoretical Physics, National Central University, Chungli 320, Taiwan}

\author{Dmitri V. Gal'tsov} \email{galtsov@phys.msu.ru}
\affiliation{Department of Theoretical Physics, Moscow State University,
119899, Moscow, Russia}

\author{Dmitry G. Orlov} \email{orlov_d@mail.ru}
\affiliation{Department of Physics, National Central University,
Chungli 320, Taiwan}

\affiliation{Center for Gravitation and Fundamental Metrology,
VNIIMS, \\ 46 Ozyornaya St., Moscow 119361, Russia}

\date{\today}

\begin{abstract}
We investigate extremal dyon black holes in the
Einstein-Maxwell-dilaton (EMD) theory with higher curvature
corrections in the form of the Gauss-Bonnet density coupled to the
dilaton. In the same theory without the Gauss-Bonnet term the
extremal dyon solutions exist only for discrete values of the
dilaton coupling constant $a$. We show that the Gauss-Bonnet term
acts as a dyon hair tonic enlarging the allowed values of $a$ to
continuous domains in the plane  $(a,\,q_m)$ the second parameter
being the magnetic charge. In the limit of the vanishing curvature
coupling (a large magnetic charge) the dyon solutions obtained tend
to the Reissner-Nordstr\"om solution but not to the extremal dyons
of the EMD theory. Both solutions have the same values of the
horizon radius as a function of charges. The entropy of new dyonic
black holes interpolates between the Bekenstein-Hawking value in the
limit of the large magnetic charge (equivalent to the vanishing
Gauss-Bonnet coupling) and twice this value for the vanishing
magnetic charge. Although an expression for the entropy can be
obtained analytically using purely local near-horizon solutions, its
interpretation as the black hole entropy is legitimate only once the
global black hole solution is known to exist, and we obtain
numerically the corresponding conditions on the parameters. Thus, a
purely local analysis is insufficient to fully understand the
entropy of the curvature corrected black holes. We also find dyon
solutions which are not asymptotically flat, but approach the linear
dilaton background at infinity. They describe magnetic black holes
on the electric linear dilaton background.
\end{abstract}


\maketitle

\section{Introduction}
In a recent paper~\cite{Chen:2006ge} we started an investigation of
extremal black holes in the four-dimensional
Einstein-Maxwell-dilaton-Gauss-Bonnet model motivated by an interest
to microscopic string calculations of the black hole entropy (for a
review see~\cite{deWit:2005ya, Mohaupt:2005jd}). In the theories
with higher curvature corrections, the entropy deviates from the
Bekenstein-Hawking value and can be calculated using Wald's
formalism~\cite{Wald:1993nt, Jacobson:1993vj, Iyer:1994ys,
Jacobson:1994qe, Myers:1998gt, Callan:1988hs}. Remarkably, it still
exhibits an agreement with the string theory predictions at the
corresponding level, both in the BPS~\cite{Behrndt:1998eq,
LopesCardoso:1998wt, LopesCardoso:1999cv, LopesCardoso:1999ur,
LopesCardoso:1999xn, Mohaupt:2000mj, LopesCardoso:2000qm,
LopesCardoso:2000fp, Dabholkar:2004yr, Dabholkar:2004dq, Sen:2004dp,
Hubeny:2004ji, Bak:2005mt} and non-BPS~\cite{Goldstein:2005hq,
Kallosh:2005ax, Tripathy:2005qp, Giryavets:2005nf, Goldstein:2005rr,
Kallosh:2006bt, Kallosh:2006bx, Prester:2005qs, Cvitan:2007pk,
Cvitan:2007hu, Cvitan:2007en, Alishahiha:2006ke, Sinha:2006yy,
Chandrasekhar:2006kx, Parvizi:2006uz, Sahoo:2006rp,
Astefanesei:2006sy} cases. In some supersymmetric models with higher
curvature terms exact classical solutions for static black holes
were obtained~\cite{Dabholkar:2004yr, Dabholkar:2004dq, Bak:2005mt}.
Moreover, as was argued by Sen~\cite{Sen:2005wa, Sen:2005iz,
Sen:2007qy}, the knowledge of the global black hole solutions is not
necessary in order to compare classical and quantum results for the
entropy: in the classical theory the entropy can be computed locally
using the entropy function approach based on the attractor property
typical for supergravity black holes~\cite{Goldstein:2005hq,
Kallosh:2005ax, Tripathy:2005qp, Giryavets:2005nf, Goldstein:2005rr,
Kallosh:2006bt}. However the question remains, whether the global
black holes corresponding to local solutions used to construct the
entropy function do really exist. Generically, the existence of
local solutions exhibiting the event horizons does not guarantee
that they describe black holes which must be regular outside the
horizon and asymptotically flat. Even though formally the entropy
can be obtained from local considerations, the parameters involved
may be subject to restrictions which are revealed only when we try
to extend the local solutions to infinity. This issue is addressed
in the present paper for extremal dyon black holes with the
$AdS_2\times S^2$ horizon in the Gauss-Bonnet gravity.

The extremal dilatonic black
hole~\cite{Gibbons:1982ih,Gibbons:1985ac,Gibbons:1987ps,Garfinkle:1990qj}
is a particularly interesting model associated with the heterotic
string. In the Einstein-Maxwell-dilaton (EMD) theory without
curvature corrections such a solution have a singular horizon of
zero radius, the corresponding Bekenstein-Hawking entropy being
zero. Typical higher curvature correction to this theory is given by
the Gauss-Bonnet  density coupled to the dilaton (later on referred
as the Einstein-dilaton-Gauss-Bonnet (EDGB) model). In this model
the local solutions with the $AdS_2\times S^2$ (Reissner-Nordstr\"om
type) horizons of finite radius can be found. Presumably they should
describe the extremal black holes possessing non-zero entropy.
In~\cite{Chen:2006ge} we have explored whether the local solutions
constructed as series expansions in the vicinity of the event
horizon can be extended to infinity as asymptotically flat black
holes. It was shown that this is possible only within some bounded
region in the space of parameters. Actually, the model contains two
parameters which are worth to be considered as independent in the
classical theory: the Gauss-Bonnet coupling constant $\alpha$
weighting the contribution of the Gauss-Bonnet term, and the dilaton
coupling constant $a$. In the case of the purely electric
configurations, the local solutions around the $AdS_2\times S^2$
horizon exist only for non-zero $\alpha$, since no dilatonic purely
electric (or purely magnetic) black holes with the $AdS_2\times S^2$
horizon are possible in the Einstein-Maxwell-dilaton (EMD) theory
without curvature corrections. Thus the only continuously varying
parameter is the dilaton coupling constant $a$. If $a=0$, the model
reduces to the Einstein-Maxwell theory, in which the desired
extremal solution do exist (the extremal Reissner-Nordstr\"om
solution). We have studied whether similar solutions exist in the
curvature-corrected theory for non-zero $a$, aiming to investigate
the case $a=1$ relevant to the compactified heterotic string. It
turned out that the extremal black hole solutions exist only in a
bounded region $0\leq a\le a_{\rm cr}$ where the critical value
$a_{\rm cr}$ is close to $1/2$. For greater $a$, the local solutions
exhibiting the $AdS_2\times S^2$ horizons can not be extended to
infinity as regular black holes, but develop singularities outside
the horizon. The threshold value of $a$ might serve an indication
that the corresponding string configuration experiences some
qualitative change of state like the black hole-string
transition~\cite{Cornalba:2006hc}.

The case of the purely electric extremal black hole in the EDGB
model is special in the sense that no similar solutions exist in the
EMD theory without curvature corrections (unless $a=0$), so there is
no smooth transition to the non-corrected theory. The situation
becomes more flexible if we allow  for both electric and magnetic
charges to be present, since extremal dyons with the $AdS_2\times
S^2$ horizon do exist in the EMD theory. As we will see, such
solutions are possible for discrete values of the dilaton coupling
constant: $a^2=1$ (the heterotic case), $a^2=3$ (the Kaluza-Klein
case) and some sequence of other integer $a_i^2$ (coinciding with
the sequence found in~\cite{Poletti:1995yq} for existence of
non-extremal black holes with two horizons). Therefore in the
curvature-corrected theory we have an infinite sequence of starting
points  for the dilaton coupling constant, not just the trivial
point $a=0$. So one can expect to have much larger domain of
existence of extremal charged dyonic black holes in the EDGB model
than in the purely electric case.

We will be interested here uniquely by the extremal black holes with
the degenerate event horizon. Non-degenerate black holes in the same
theory were extensively studied in the past both
perturbatively~\cite{Mignemi:1992nt, Mignemi:1993ce} and
numerically~\cite{Kanti:1995vq, Torii:1996yi, Alexeev:1996vs,
Alexeev:1997ua, Guo:2008hf}. More recently global properties of EDGB
black hole solutions were studied using the dynamical system
approach~\cite{Melis:2005xt, Melis:2005ji, Mignemi:2006ut,
Melis:2006fj}. Stability issues were discussed
in~\cite{Torii:1998gm,Dotti:2004sh,Dotti:2005sq,Gleiser:2005ra,Moura:2006pz}.
In these papers the existence of both neutral and charged
asymptotically flat solutions with a non-degenerate event horizon
and without naked singularities was established. These solutions
have the Schwarzschild type event horizon and they do not possess
the extremal limits. The solution with the degenerate event horizon
thus form a separate branch of EDGB black holes which was not
studied before.

The existence of non-degenerate dyonic black holes in the EMD theory
without curvature corrections was studied in detail
in~\cite{Poletti:1995yq}. It was found that such solutions
generically exhibit one (non-degenerate) horizon of the
Schwarzschild type, but for some discrete values of the dilaton
coupling constant there are solutions with two horizons of the
Reissner-Nordstr\"om type. In this latter case the limiting extremal
solutions turn out to be possible. We will show that adding the
Gauss-Bonnet term to the EMD action acts as a hair tonic for
extremal dyons, allowing for continuously varying dilaton couplings.
Still, the allowed domain is bounded (in somewhat irregular way), so
the threshold behavior observed in~\cite{Chen:2006ge} persists in
the dyon case too. The solutions corresponding to the values of
parameters approaching the boundary of the allowed domain exhibit
interesting saturation properties similar to the BPS bounds in the
EMD theory without curvature corrections.

The plan of the paper is as follows. The Sec.~2 contains general
definitions and investigation of the hidden symmetries of the
reduced theory. It is shown that the one-dimensional theory
possesses the two-parametric off-shell and the three-parametric
on-shell symmetry groups. These symmetries serve as a convenient
tool allowing to describe the space of solution semi-analytically,
in spite of the fact that the equations of motion do not have
analytic solutions. In Sec.~3 we investigate local series solutions
around the assumed $AdS_2\times S^2$ horizon. We show that for fixed
values of the electric and magnetic charges the local solutions
contain only one free parameter, contrary to two parameters in the
corresponding theory without curvature corrections. Examining the
higher order coefficients we find the family of curves in the
parameter plane of $a$ and $q_m$ (the magnetic charge) on which the
coefficients of the local solution are singular. These curves are
parameterized by the same sequence of integers $a_i^2$ which
correspond to existence of dyons in the uncorrected theory.  In this
section we also calculate the entropy and show that it interpolates
between the Bekenstein-Hawking value $A/4$ in the case of vanishing
Gauss-Bonnet coupling  (or large magnetic charge) and  twice the
Bekenstein-Hawking large value $A/2$ for purely electric solutions,
as found in~\cite{Chen:2006ge}. The Sec.~4 is devoted to asymptotic
solutions. We find that, similarly to the uncorrected EMD theory,
there are two physically interesting asymptotic patterns
corresponding either to usual asymptotically flat black holes, or to
black holes on the linear dilaton background~\cite{Clement:2002mb}.
Numerical results are presented in the Sec.5. Using the on shell
symmetries we can express the ratios of physical parameters as
functions of the dilaton coupling constant only, which clarifies the
properties of the whole family of dyonic solutions obtained. We also
demonstrate their asymptotic BPS-type behavior on the boundary of
the allowed domain of parameters. The parameter region for the
asymptotically LDB solutions is shown to locate on the other side of
the limiting singular curve in parameter plane. In the Appendix A we
give some details concerning the EMD dyons and show how analytic
solutions known for two lower values of the discrete dilaton
coupling sequence can be obtained by summing up the local series
solutions valid in the vicinity of the horizon. In the Appendix B
some higher order coefficient of the local solution in the EDGB
model are listed which are necessary in deriving the existence of
limiting curves in the parameter space.

\section{General setting}
We consider the four-dimensional dilatonic Gauss-Bonnet theory
(EDGB) which is the Einstein-Maxwell-dilaton theory (EMD) with an
arbitrary dilaton coupling constant $a$ modified by the Gauss-Bonnet
(GB) term:
\begin{equation}\label{reducedS}
S = \frac1{16 \pi} \int  \left\{ R - 2 \partial_\mu \phi \partial^\mu \phi
- {\rm e}^{2 a \phi} \left( F^2 - \alpha {\cal L}_{\rm GB} \right) \right\}
\sqrt{-g} \, d^4x,
\end{equation}
where ${\cal L}_{\rm GB}$ is the Gauss-Bonnet density
\begin{equation}
{\cal L}_{\rm GB} = R^2 - 4 R_{\mu\nu} R^{\mu\nu} + R_{\alpha\beta\mu\nu}
R^{\alpha\beta\mu\nu}.
\end{equation}
This action contains two parameters (we use the units $G = c = 1$):
the dilaton coupling constant $a$ and the Gauss-Bonnet coupling
constant $\alpha$. We assume $a \geq 0, \; \alpha \geq 0$, solutions
for negative $a$ can be obtained changing the sign of the dilaton.
Note that in this action the Maxwell term is not multiplied by
$\alpha$ to facilitate decoupling of the Gauss-Bonnet term from the
EMD action.

Consider the static spherically symmetric metrics~\footnote{A
detailed discussion of the gauge fixing can be found
in~\cite{Chen:2006ge}. In this paper we adopt the gauge $g_{tt} =
1/g_{rr}$.}, parameterized by two functions $w(r)$ and $\rho(r)$:
\begin{equation}
ds^2 = - w(r) dt^2 + \frac{dr^2}{w(r)} + \rho^2(r) d\Omega_2^2,
\end{equation}
the scalar curvature and the Gauss-Bonnet density then read
\begin{eqnarray}
R &=& \frac1{\rho^2} \left[ - (4 w \rho \rho' + w' \rho^2)' +
2 \rho' (w \rho)' + 2 \right],
\\
{\cal L}_{\rm GB} &=& \frac4{\rho^2} [ w' (w \rho'^2 - 1) ]'.
\end{eqnarray}
The corresponding ansatz for the Maxwell one-form is
\begin{equation}
A = - f(r) \, dt - q_m \cos\theta \, d\varphi,
\end{equation}
where $f(r)$ is the electrostatic potential and $q_m$ is the
magnetic charge. Note that the Gauss-Bonnet term breaks the discrete
S-duality of the EMD theory without curvature corrections which  is
described by the transformation
\begin{equation}
g_{\mu\nu} \to g_{\mu\nu}, \qquad F \to {\rm e}^{-2a\phi} {}^* F,
\qquad \phi \to - \phi,
\end{equation}
where $F = dA$. It can be expected therefore that the properties of
electric or magnetic black holes in this theory will be be
essentially different.

\subsection{Reduced action and  equations of motion}
The one-dimensional Lagrangian associated with the ansatz is
obtained by dropping the total derivative in the dimensionally
reduced action:
\begin{equation}
L = \frac12  [\rho' (w \rho)' + 1] - 2 \alpha a w' (w \rho'^2 - 1)
\phi' {\rm e}^{2 a \phi} - \frac12 w \rho^2 \phi'^2 + \frac12 \rho^2
f'^2 {\rm e}^{2 a \phi} - \frac12 \frac{q_m^2}{\rho^2} {\rm e}^{2 a \phi}.
\end{equation}
The corresponding equations of motion read:
\begin{eqnarray}
4 \alpha a \left[ (w \rho'^2 - 1) \phi' {\rm e}^{2 a \phi} \right]' -
4 \alpha a w' \rho'^2 \phi' {\rm e}^{2 a \phi} - \rho \rho'' - \rho^2 \phi'^2
&=& 0, \label{Eqw}
\\
4 \alpha a (w w' \rho' \phi' {\rm e}^{2 a \phi})' - \frac12 w'' \rho - (w \rho')'
- w \rho \phi'^2 + \rho f'^2 {\rm e}^{2 a \phi} + \frac{q_m^2}{\rho^3}
{\rm e}^{2 a \phi} &=& 0, \label{Eqrho}
\\
(w \rho^2 \phi')' + 2 \alpha a [ w' (w \rho'^2 - 1) ]' {\rm e}^{2 a \phi}
+ a \rho^2 f'^2 {\rm e}^{2 a \phi} - a \frac{q_m^2}{\rho^2} {\rm e}^{2 a \phi}
&=& 0, \label{Eqphi}
\\
\left( \rho^2 f' {\rm e}^{2 a \phi} \right)' &=& 0. \label{Eqf}
\end{eqnarray}
Integrating once the last equation for the form field (\ref{Eqf}),
\begin{equation}\label{Solf}
f'(r) = q_e \rho^{-2} {\rm e}^{- 2 a \phi},
\end{equation}
where $q_e$ is the electric charge parameter, one can then insert
(\ref{Solf}) into the previous equations to obtain the six-order
system consisting of three differential equations of the second
order with the electric and magnetic charges $q_e, q_m$ entering as
fixed parameters.

\subsection{Symmetries and conserved quantities}
The action (\ref{reducedS}) is invariant under the following two-parametric
group of global transformations:
\begin{equation}\label{symL}
w \to w \, {\rm e}^{-2 \delta}, \quad \rho \to \rho \, {\rm e}^\delta,
\quad \phi \to \phi + \frac{\delta}{a}, \quad f \to f \, {\rm e}^{-2 \delta};
 \qquad r \to r + \nu,
\end{equation}
which generate two conserved Noether currents
\begin{equation}
J_g := \left( \frac{\partial L}{\partial \Phi'^A} \Phi'^A - L \right)
\partial_g r \bigg|_{g=0} - \frac{\partial L}{\partial \Phi'^A} \,
\partial_g \Phi^A \bigg|_{g=0}, \qquad \partial_r J_g = 0,
\end{equation}
where $\Phi^A$ stands for the set $w, \rho, \phi, f$, and $g =
\delta,\, \nu$ being the transformation parameters. The conserved
quantity corresponding to $\nu$ is the Hamiltonian
\begin{equation}\label{J1}
H = \frac12 [\rho' (w \rho)' - 1] - 2 \alpha a w' (3 w \rho'^2 - 1)
\phi' {\rm e}^{2 a \phi} - \frac12 w \rho^2 \phi'^2 + \frac12 \rho^2
f'^2 {\rm e}^{2 a \phi} + \frac12 \frac{q_m^2}{\rho^2} {\rm e}^{2 a \phi}.
\end{equation}
This quantity must vanish on shell for diffeomorphism invariant
theories, so $H = 0$. The Noether current corresponding to $\delta$
leads to the conservation equation $\partial_r J_\delta = 0$ with
the current
\begin{equation}
J_\delta = \frac{w \rho^2 \phi'}a  - \frac{w' \rho^2}2 + 2 q_e f +
2 \alpha \left[ (w \rho'^2 - 1) (w' - 2 a w \phi') + 2 a w w' \rho \rho' \phi'
\right] {\rm e}^{2 a \phi},
\end{equation}
which is an Abelian counterpart of the integral given
in~\cite{Donets:1995ya}.

The integrals of motion allow one to reduce the order of the system
by two leading to the forth order system. Moreover, for $q_e = 0$
one can further reduce the order to three introducing the new
variables
\begin{equation}
w \to \exp(w), \qquad \rho \to \exp \left(\rho - \frac{w}2 \right),
\qquad \phi \to \phi - \frac1{2 a} w.
\end{equation}
Using them we can exclude from the system $w$, while $w'$ and $w''$
still remain. For numerical integration we will still use the
initial six-dimensional system, applying the integrals of motion to
control accuracy of the calculation.

The symmetry group is enlarged on-shell on the space of the
solutions of the equations of motion. It can be easily seen that the
solution space is invariant under a {\sl three-parameteric} group of
global transformations which consists in rescaling of the electric
charge
\begin{equation}\label{tqe}
q_e \to q_e \, {\rm e}^{2 \delta}, \qquad q_m \to q_m,
\end{equation}
(leaving the magnetic charge invariant), rescaling and shift of an
independent variable
\begin{equation}\label{trr}
r \to r \, {\rm e}^{\frac{\mu}2 + \delta} + \nu,
\end{equation}
and the following transformation of the field functions:
\begin{equation} \label{symsol}
w \to w \, {\rm e}^\mu, \qquad \rho \to \rho \, {\rm e}^\delta, \qquad \phi
\to \phi
+ \frac{\delta}{a}, \qquad f \to f \, {\rm e}^{\frac{\mu}2 - \delta}.
\end{equation}
These do not leave the action (\ref{reducedS}) invariant, unless a
condition
\begin{equation}
\mu = - 2 \delta
\end{equation}
is imposed in which case we go back to the transformations
(\ref{symL}). However, dealing with the solutions,  we can assume
both parameters $\mu$ and $\delta$ to be  independent.

\section{Local solutions with $AdS_2 \times S^2$ horizon}
We are looking for extreme black holes carrying both electric and
magnetic charges in the EDGB theory for which the metric function
$w(r)$ has double zero at some point $r = r_H$ (the event horizon)
and is non-singular for $r > r_H$. Already in the EMD theory without
curvature corrections  the analytical dyonic solutions are known
only for two special values of the dilaton coupling, namely $a^2 =
1$ and $a^2 = 3$ (for a more detailed discussion see Appendix A)
while for generic values of $a$  only a numerical analysis is
possible. So we can hope to solve the problem of dyonic black holes
in the EDGB theory only numerically. Meanwhile, already from a local
analytical analysis of the solution near the event horizon we obtain
important restrictions on the parameters.

\subsection{Horizon expansions}
Local solutions in the vicinity of the event horizon $r = r_H$ can
be constructed  expanding them in terms  of the deviation $x = r - r_H$:
\begin{equation}
w(r) = \sum_{k=2}^\infty w_k x^k, \qquad \rho(r) = \sum_{k=0}^\infty \rho_k x^k,
\qquad P(r) := {\rm e}^{2 a \phi(r)} = \sum_{k=0}^\infty P_k x^k.
\end{equation}
According to the assumption of extremality, the function $w(r)$
starts with the quadratic term: vanishing of $w_0$ means that $r =
r_H$ is a horizon and vanishing of $w_1$ means that the horizon is
degenerate. Such an expansion contains only one free parameter $P_1$
for fixed values of charges:
\begin{eqnarray} \label{locdeg}
w(r) &=& \frac{x^2}{\rho_0^2} - \frac{P_1}{6 \alpha a^2 \rho_0^4}
\left[ 3 (a^2 - 1) q_m^4 + 6 \alpha (3 a^2 - 2) q_m^2 + 4 \alpha^2 (5 a^2 - 3)
\right] x^3 + O(x^4),
\nonumber\\
\rho(r) &=& \rho_0 + \frac{P_1}{4 \alpha a^2 \rho_0} \left[ (a^2 - 1) q_m^4
+ 2 \alpha (3 a^2 - 2) q_m^2 + 4 \alpha^2  (a^2 - 1) \right] x + O(x^2),
\nonumber\\
P(r) &=& \frac{\rho_0^2}{2 (2 \alpha + q_m^2)} + P_1 x + O(x^2).
\end{eqnarray}
The physical value of the horizon radius  $\rho_0$ is not a free
parameter here: it is fixed by the charges as follows
\begin{equation}\label{rho0}
\rho_0^2 = \frac{2 q_e (2\alpha + q_m^2) }{\sqrt{4 \alpha + q_m^2}}.
\end{equation}
Note that the dilaton coupling constant enters the expansions only
through $a^2$, so the space of solutions is symmetric under $a \to -
a,\, \phi \to - \phi$ (we assume $a \ge 0$). When $a\to 0$, some of
the expansion coefficients diverge unless $P_1 = 0$. If $P_1 = 0$,
the higher order coefficients in the dilaton expansion also vanish
and the dilaton is constant. Then the Gauss-Bonnet terms in the
action becomes a total derivative, so the theory reduces to the
Einstein-Maxwell one. One can then show that the series expansions
for $w$ and $\rho$ then combine indeed into the extremal
Reissner-Nordstr\"om solution.

Higher order expansion coefficients (see Appendix B) contain in
denominators a sequence of the following combinations of the
parameters
\begin{equation}\label{ajqm}
\Upsilon_i (a,\,q_m):= (a^2 - a_i^2) q_m^4 + 2 \alpha (3 a^2 - 2
a_i^2) q_m^2 + 4 \alpha^2 [ a^2 (a_i^2 + 2) - a_i^2 ], \qquad i \ge
2,
\end{equation}
where $a_i^2$ are the integers:
\begin{equation}\label{ai}
a_i^2 = 1 + 2 + \cdots + i = \frac{i (i + 1)}2.
\end{equation}
Therefore the expansions do not exist for the values of $a$ and
$q_m$ satisfying the equations
\begin{equation}
\Upsilon_i = 0, \qquad i \ge 2,
\end{equation}
which define the sequence of limiting curves in the parameter plane
$a,\, q_m$. The only possibility to avoid the divergences of the
expansion coefficients would be to choose the parameter $P_1$ as the
product $P_1 = c \prod_{i=2}^\infty \Upsilon_i$ with finite $c$.
However, in this case there will be a turning point outside the
horizon in which $d \rho/d r = 0$, so that the solution does not
extend to infinity but ends up in a singularity.

Remarkably, the sequence of integers (\ref{ai}) coincides with that
found by Poletti et al.~\cite{Poletti:1995yq} as condition of
existence of  asymptotically flat dyons with two (non-degenerate)
horizons in the EMD theory without curvature corrections. In our
case this sequence enters in the definition of the set of functions
$\Upsilon_i(a, q_m)$, whose vanishing marks non-existence of the
extremal local solution. In the limit $q_m \to \infty$ which is
equivalent to $\alpha \to 0$ (decoupling of the Gauss-Bonnet term)
the solution of the equation $\Upsilon_i(a, q_m) = 0$ is just $a =
a_i$. Specializing the path in the $a,\, q_m$ plane one can in
principle make higher order coefficients non-singular, but the
procedure is somewhat subtle (a more detailed discussion will
follow).

Another characteristic curve in the parameter plane $a,\, q_m$  is
defined by vanishing of the linear term in the expansion of the
radial function $\rho(r)$:
\begin{equation}
\Upsilon_1 := (a^2 - 1) q_m^4 + 2 \alpha (3 a^2 - 2) q_m^2 + 4
\alpha^2 (a^2 - 1)=0.
\end{equation}
This signals a potential singularity of the corresponding global
solution (the horizon itself is a turning point for the radial
variable). Note that the curve $a(q_m)$ obtained as the solution of
the equation $\Upsilon_1 = 0$ reaches the value $a = 1$ for $q_m=0$,
and $q_m \to \infty$, and it has a local minimum $a^2 = 4/5$ for
$q_m = \sqrt{2 \alpha}$. It is worth noting that the expression for
$\Upsilon_1$ does not follow the general formula (\ref{ajqm}) for
$\Upsilon_i$ (valid for $i\geq 2$).

Therefore, the family of curves $\Upsilon_i = 0$ divide the
two-dimensional parameter space of $a$ (vertical axis) and $q_m$
(horizontal axis) of the global solutions for a fixed $\alpha$ into
the disconnected regions. This situation will be described in detail
in Sec. IV.

The values of the integrals of motion corresponding to the series
expansions (\ref{locdeg}) are
\begin{equation}
H = \frac1{2 \rho_0^2} \left[ q_m^2 P_0 + q_e^2 P_0^{-1} - \rho_0^2
\right]=0, \qquad J_\delta = 2 \, q_e \, f_0,
\end{equation}
where $P_0$ is given explicitly in (\ref{locdeg}) and $f_0$ is the
value of the electrostatic potential on the horizon.

The unique free parameter $P_1$ in the near horizon expansions will
be fixed by asymptotic flatness. This means that the extremal dyonic
solutions are completely characterized by the charges. It is
convenient to absorb the Gauss-Bonnet coupling constant
$\alpha$ into the redefinition of the parameters
$q_e = \hat q_e/ \sqrt\alpha, \; q_m = \sqrt\alpha \,
\hat q_m$ and $P_1 = \hat P_1/\alpha$. Then the near horizon
expansions then will read:
\begin{eqnarray}\label{locdeghat}
w(r) &=& \frac{x^2}{\rho_0^2} - \frac{\hat P_1}{6 a^2 \rho_0^4} \left[ 3 (a^2 - 1)
\hat q_m^4 + 6 (3 a^2 - 2) \hat q_m^2 + 4 (5 a^2 - 3) \right] x^3 + O(x^4),
\nonumber\\
\rho(r) &=& \rho_0 + \frac{\hat P_1}{4 a^2 \rho_0} \left[ (a^2 -1) \hat q_m^4 +
2 (3 a^2 - 2) \hat q_m^2 + 4 (a^2 - 1) \right] x + O(x^2),
\nonumber\\
\hat P(r)&=&  \alpha \mathrm{e}^{2 a \phi} = \frac{\rho_0^2}{2 (2 +
\hat q_m^2)} +
 \hat P_1 \, x + O(x^2),
\end{eqnarray}
and the relation (\ref{rho0}) becomes
\begin{equation}
\rho_0^2 = \frac{2 \hat q_e (2 + \hat q_m^2)}{\sqrt{4 + \hat q_m^2}}.
\end{equation}
It is important to determine the correct sign of $\hat P_1$. To be
able to interpret the region $r > r_H$ as an exterior of the black
hole, one has to ensure positiveness of the derivative $\rho'$ at
the horizon. From the near horizon expansion of $\rho$ one finds
\begin{equation}
\rho'|_{x=0} = \frac{\hat P_1}{4 a^2 \rho_0} \hat \Upsilon_1 > 0,
\qquad \hat \Upsilon_1 := (a^2 -1) \hat q_m^4 + 2 (3 a^2 - 2) \hat
q_m^2 + 4 (a^2 - 1).
\end{equation}
Thus, we should take positive $\hat P_1$ for $\hat \Upsilon_1 > 0$
and negative $\hat P_1$ for $\hat \Upsilon_1 < 0$. Introducing the
sign parameter $ \varsigma = \frac{\hat P_1}{|\hat P_1|}, $ we find
therefore:
\begin{equation}
\varsigma  = \frac{\hat \Upsilon_1}{|\hat \Upsilon_1|}.
\end{equation}
Another useful redefinition is based on the observation that
$\rho_0$ and $P_1$ enter the expansions in the combination $b =
|\hat P_1|/(a^2 \rho_0^2)$. Consider now the transformations of the
expansion parameters under the symmetries of the solution space
(\ref{trr}, \ref{symsol}). It is easy to see that the full set of
local solutions can be generated from one particular solution with
$\rho_0 = 1, \, b = 1$, which we will call the normalized local
solution, by the symmetry transformations with $\delta = - \ln
\rho_0$ and $\mu = 2 \ln (b \rho_0)$. The normalized local solution
does not contain free parameters (for fixed $q_m$):
\begin{eqnarray}\label{locdegnor}
w(r) &=& x^2 - \varsigma \frac16 \left[ 3 (a^2 - 1) \hat q_m^4 + 6 (3 a^2 - 2)
\hat q_m^2 + 4 (5 a^2 - 3) \right] x^3 + O(x^4),
\nonumber\\
\rho(r) &=& 1 + \varsigma \frac14 \left[ (a^2 -1) \hat q_m^4 + 2 (3 a^2 - 2)
\hat q_m^2 + 4 (a^2 - 1) \right] x + O(x^2),
\nonumber\\
\hat P(r) &=& \frac1{2 (2 + \hat q_m^2)} + \varsigma a^2 x + O(x^2).
\end{eqnarray}
Note the presence of the sign function $\varsigma$ in the odd power
terms. The electric charge corresponding to the normalized local
solution is given by
\begin{equation}
\hat q_e = \frac{\sqrt{4 + \hat q_m^2}}{4 + 2 \hat q_m^2}.
\end{equation}

\subsection{Reissner-Nordstr\"om limit}
One case in which the extremal Reissner-Nordstr\"om solution is
valid is the already mentioned limit $a \to 0$. However, there is
another limit in which our local solutions make contact with the
Reissner-Nordstr\"om solution: $\hat q_m \to \infty$. This can be
implemented by taking either $q_m \to \infty$ or $\alpha \to 0$. In
this limit, the near horizon expansions (\ref{locdeghat}) reduce to
\begin{eqnarray}\label{locdeghatqm}
w(r) &=& \frac{x^2}{\rho_0^2} \left[ 1 - 2 F_1 x + 3 F_1^2 x^2 - 4
F_1^3 x^3 + O(x^4) \right],
\nonumber\\
\rho(r) &=& \rho_0 \left[ 1 + F_1 x + \frac{a^2 - 2}{2 a^2}
\frac1{\hat q_m^4}
\left( \frac{\hat P_1 \hat q_m^4}{\rho_0^2} \right)^2 x^2 + \frac{(a^2 - 1)
(2a^4 + 19 a^2 - 36)}{24 a^4 (a^2 - 6)} \frac1{\hat q_m^4} \left( \frac{\hat P_1
\hat q_m^4}{\rho_0^2} \right)^3 x^3 + O(x^4) \right],
\nonumber\\
\hat P(r) &=& \alpha \mathrm{e}^{2 a \phi} = \frac{\rho_0^2}{2 \hat
q_m^2} + \hat P_1 \left[ x - F_1 x^2 - \frac{(a^2 - 1)^2 (a^2 +
4)}{16 a^4 (a^2 - 6)} \left( \frac{\hat P_1 \hat q_m^4}{\rho_0^2}
\right)^2 x^3 + O(x^4) \right],
\end{eqnarray}
where
\begin{equation}
F_1 = \frac{(a^2 - 1) \hat P_1 \hat q_m^4}{4 a^2 \rho_0^2}.
\end{equation}
This is valid for all values of $a$ except $a = a_i$ located on the
curves of $\Upsilon_i = 0$ when $\hat q_m \to \infty$. To ensure the
$AdS_2\times S^2$ structure of the horizon and the regularity of
expansions we assume $\rho_0$ to be finite and impose the condition
$F_1= \mbox{const} \neq 0$. Since $\rho_0^2 = 2 \hat q_e \hat q_m$,
this implies $\hat P_1 \sim \sqrt\alpha \, \hat q_m^{-3} \to 0$.
Then, the expansions simplify and admit summation to the closed
expressions
\begin{equation}
w(r) = \frac{x^2}{\rho_0^2} (1 + F_1 x)^{-2}, \qquad \rho(r) = \rho_0 (1 + F_1 x),
\qquad \hat P(r) = \hat P_0.
\end{equation}
Now, the asymptotic flatness means $F_1 = 1 / \rho_0$, and finally
assuming $r_H = \rho_0$ we will have $x = r - \rho_0$ which lead to
the Reissner-Nordstr\"om dyonic black hole
\begin{equation}
w(r) = \left( 1 - \frac{\rho_0}{r} \right)^2, \qquad \rho(r) = r, \qquad \hat P(r)
= \hat P_0 = \alpha \mathrm{e}^{a \phi_h},
\end{equation}
with the electric and magnetic charge parameters $\hat q_e =
\sqrt{\frac{\hat P_0}2} \rho_0$ and $\hat q_m = \rho_0 / \sqrt{2
\hat P_0}$ (corresponding to equal dilaton-rescaled charges $Q_e =
Q_m = \rho_0/\sqrt2$).

Note again that there are two different ways to implement the limit
$\hat q_m \to \infty$: either $\alpha \to 0, \hat P_0 \to 0$ and the
radius of the horizon remaining finite $\rho_0 = 2 \hat q_e \hat q_m
= 2 q_e q_m$, or $q_m \to \infty$, implying the infinitely large
radius of the horizon.

\subsection{Relation to EMD dyons}
The dyon solutions of the EMD theory without curvature corrections
for a generic $a$ exhibit one horizon and does not admit an extremal
limit. For a discrete sequence $a = a_i$ there are solutions with
two horizons which may have such limits~\cite{Poletti:1995yq}. This
sequence is the same as found above from a different reasoning. We
expect to have a relationship between our solution and those of the
Ref.~\cite{Poletti:1995yq} (in the extremal limit) when $\hat q_m
\to \infty$ (equivalent to $\alpha \to 0$). However, the situation
is somewhat subtle. Restarting with the set of equations of motion
for $\alpha = 0$ and considering the series solution near the event
horizon, we obtain in the lowest order:
\begin{equation}\label{P0rho0}
P_0 = \frac{q_e}{q_m}, \qquad \rho_0^2 = 2 q_e q_m.
\end{equation}
This corresponds to the limiting form of the coefficients in the
EDGB theory. Analyzing the higher-order equations for the expansions
coefficients we find the following. In general $\rho_1$ is a free
parameter (which has to be fixed by the asymptotic conditions) while
all $w_k$ and $\rho_k$ with $k \ge 2$ are completely determined by
the equations order by order. However, when one tries to solve the
equations with respect to $P_k$ for $k \ge 1$, an interesting
bifurcation behavior is observed. There are two possible cases:
either $a^2 = 1$ (then $P_1$ is a free parameter) or $P_1 = 0$ (we
leave aside the special case $a = 0$.) In the first case, all higher
order $P_k$ are fixed by the equations. In the second case we
observe another bifurcation: either $a^2 = 3$ (and $P_2$ is then
free) or $P_2 = 0$. Again, in the first case $P(r)$ is fixed by
$P_2$, but in the second case  we have a further bifurcation: either
$a^2 = 6$ or $P_3 = 0$. The analogous bifurcations exist in any
order. This branching procedure reproduces the value $a = a_i$ at
$i$-th step. This indicates that the extremal dyonic black holes can
exist only for this discrete sequence.

Thus, for any $i \ge 1$, there are two independent parameters
$\rho_1, P_i$ in the local solution ($\rho_0, P_0$ being related to
the charge parameters), and the expansion coefficients $\rho_j, j =
2, \dots, 2 i - 1$ and $P_j, j = 1, \dots, i - 1$ are {\em all
zero}. Moreover, the expansion of $w(r)$ differs from the
corresponding expansion of the Reissner-Nordstr\"om solution ($a =
0$)
\begin{equation}
w_{\rm RN}(r) = \frac{x^2}{(\rho_0+\rho_1 x)^2}=\frac1{\rho_0^2} \,
x^2 - \frac{2 \rho_1}{\rho_0^3} \, x^3 + \frac{3 \rho_1^2}{\rho_0^4}
\, x^4 - \frac{4 \rho_1^3}{\rho_0^5} \, x^5 + \frac{5
\rho_1^4}{\rho_0^6} \, x^6 + O(x^7),
\end{equation}
only starting with the term $w_{2i + 2}$.

In view of such a behavior, to reach the series expansions arising
in the EMD theory, which could be expected to arise in the limit
$\alpha = 0$ of the EDGB model is somewhat problematic. First, in
the EDGB case there is only one free parameter, $P_1$, while there
are {\em two} parameters, $\rho_1$ and $P_i$, in the EMD theory. So
the solutions emerging in the limit $\hat q_m \to \infty$, if exist,
can contact the corresponding solutions in the EMD theory only for a
special value of $\rho_1$. Secondly, the limit $a\to a_i, \; \hat
q_m \to \infty$ in the parameter space depends on the direction
chosen. In particular, one gets essentially different results taking
first $a \to a_i$ and then $\hat q_m \to \infty$, or first $\hat q_m
\to \infty$ and then $a \to a_i$. And both these two do not seem to
give the result of the EMD theory.

We refer the reader to Appendix A for details concerning the cases
of lower values of the sequence $a_i$ for which a closed form
summation is possible.

\subsection{Entropy and temperature}
Following the Sen's entropy function approach, the entropy of
extremal dyonic black holes can be straightly calculated
\begin{equation}
S = 2 \pi q_e \sqrt{q_m^2 + 4 \alpha} = \pi \rho_0^2 + \frac{2 \pi
\alpha \rho_0^2}{2 \alpha + q_m^2}.
\end{equation}
Technical details are similar to those in the pure electric case
treated in~\cite{Chen:2006ge}. There are two interesting limits of
the above expression. Firstly, the Bekenstein-Hawking entropy-area
relation, $S = A/4, \; A = 4 \pi \rho_0^2$, is recovered when
$\alpha = 0$ or $q_m \to \infty$. Secondly, if the magnetic charge
parameter $q_m$ is vanishing, we recover the result obtained for
the pure electric case~\cite{Chen:2006ge, Chen:2008px}, namely, the
double Bekenstein-Hawking value. Such entropy enhancement for small
black holes was discussed in Ref.~\cite{Cai:2007cz}. However, for a
generic extremal dyonic solution, the black hole entropy can not be
completely expressed in terms of its horizon area.

It is worth noting that although the value of the entropy can be
calculated using only the local solution valid in the vicinity of
the $AdS_2\times S^2$ event horizon, its interpretation as the
entropy of a black hole presumes an existence of the global
solutions extending to infinity. We will see later on that this
imposes certain restrictions on the values of the magnetic charge
and the dilaton coupling constant $a$. Purely local analysis is
therefore insufficient for drawing conclusions about the
correspondence between the string and geometric values of the
entropy.

The temperature of the extremal EDGB black hole is zero, as for the
extremal solution without the Gauss-Bonnet term:
\begin{equation}
T = \frac1{2 \pi} \left. \left( \sqrt{g^{rr}} \; \frac{\partial
\sqrt{g_{tt}}} {\partial r} \right) \right|_{r = r_H} = \frac1{2 \pi
\rho_0^2}(r - r_H)|_{r = r_H} = 0.
\end{equation}

\section{Asymptotic behavior}
Another region where local solutions can be constructed analytically
is the asymptotic zone $r \to \infty$. Similarly to the case on the
uncorrected EMD theory, we find that two type black holes can exist:
usual asymptotically flat solutions, and black holes on the linear
dilaton background~\cite{Clement:2002mb}. In the latter case the
dilaton diverges at infinity (linearly for a proper choice of the
radial coordinate), but the ADM mass of the black hole itself is
finite, so that the solution can be interpreted as the black hole on
the linear dilaton background.

\subsection{Asymptotically flat solutions}
Looking for asymptotically flat global solutions we have
to ensure $w \to 1,\; \rho/r \to 1,\; \phi \to \phi_\infty$
(constant) as $r \to \infty$. The subleading terms should be
expandable in the power series of $1/r$. The asymptotic solution
with these properties contains five parameters: the ADM mass $M$,
the electric and magnetic charges $Q_e, Q_m$ (rescaled), the dilaton
charge $D$, and the asymptotic value of the dilaton $\phi_\infty$:
\begin{eqnarray}\label{asym}
w(r) &=& 1 - \frac{2 M}r + \frac{\alpha Q_e^2 + \alpha^{-1} Q_m^2}{r^2} +
O(r^{-3}),
\nonumber\\
\rho(r) &=& r - \frac{D^2}{2 r} - \frac{D (2 M D - \alpha a Q_e^2 + \alpha^{-1}
a Q_m^2)}{3 r^2} + O(r^{-3}),
\\
\phi(r) &=& \phi_\infty + \frac{D}{r} + \frac{2 D M - \alpha a Q_e^2 +
\alpha^{-1} a Q_m^2}{2 r^2} + O(r^{-3}), \nonumber
\end{eqnarray}
where
\begin{equation}
Q_e = q_e \mathrm{e}^{- a \phi_\infty}, \qquad Q_m = q_m
\mathrm{e}^{a \phi_\infty}.
\end{equation}
The dilaton charge can be also read off from an asymptotic expansion
of the dilaton exponential:
\begin{equation}
\mathrm{e}^{2 a (\phi - \phi_\infty)} = 1 + \frac{2 a D}{r} + \frac{2 a D (a D + M)
- \alpha  a^2 Q_e^2 + \alpha^{-1} a^2 Q_m^2}{r^2} + O(r^{-3}).
\end{equation}
The values of two integrals of motion in terms of the asymptotic
parameters are:
\begin{equation}\label{J1a}
H = \frac12 (w_\infty \rho'^2_\infty - 1), \qquad J_\delta = 2 \,
q_e \, f_\infty - M - \frac{D}{a}.
\end{equation}
The constant $f_\infty$ in the electric potential is usually fixed
to zero.

Behavior of the global solution which starts with the normalized
local solution (\ref{locdegnor}) at the horizon depends only on the
dilaton coupling constant $a$ and magnetic charge parameter $q_m =
\sqrt\alpha \hat q_m$. Its existence for all $a$ is not guaranteed
{\em a priori}. But, in some intervals of $a$ whose boundaries
depend on $q_m$, as can be shown numerically, there exist solutions
varying smoothly with increasing $x$ such that the function $w$ and
the derivative $\rho'$ stabilize at infinity on some constant values
$w_\infty \neq 1,\, \rho'_\infty \neq 1$. Then, using the symmetries
(\ref{trr}, \ref{symsol}) of the solution space, one can rescale the
global solution obtained to achieve the desired unit values for
these parameters. As we have argued, two parameters $\mu, \delta$
effectively replace the parameters $\rho_0, P_1$ of the
(non-normalized) local solution (\ref{locdeg}) or (\ref{locdeghat}).
So one could expect that rescaling of the solution ensuring
$w_\infty = 1,\, \rho'_\infty = 1$ would fix both quantities
$\rho_0, P_1$ on the horizon. But from the Hamiltonian constraint
equation $H = 0$ with $H$ given by the Eq. (\ref{J1a}) it is clear
that one must have $w_\infty \rho'^2_\infty = 1$ for any
asymptotically flat solution. Therefore it is enough to perform {\em
one} but not {\em two} independent rescalings in order to achieve
$w_\infty = 1,\, \rho'_\infty = 1$. Indeed, under the transformation
(\ref{trr}, \ref{symsol}), the relevant functions and parameters are
transformed as follows
\begin{equation}
w \to w \mathrm{e}^\mu, \quad \rho' \to \rho' \mathrm{e}^{\mu/2}, \quad w
\rho'^2 \to w \rho'^2; \qquad \rho_0 \to \rho_0 \mathrm{e}^\delta, \quad P_1
\to P_1 \mathrm{e}^{\delta - \mu/2}.
\end{equation}
Since the choice of $\mu, \delta$ is equivalent to the choice of
$\rho_0,\, P_1$, an invariance of the product $w \rho'^2$ means that
the solution starting on the horizon with {\em any} $\rho_0, P_1$
will reach at infinity the values $w_\infty,\, \rho'_\infty$
satisfying $w_\infty {\rho'}^2_\infty = 1$. Therefore, taking $\mu =
- \ln w_\infty$, we will achieve simultaneously $w_\infty = 1$ and
$\rho'_\infty = 1$. This means that asymptotically flat solutions
still form a two-parameter family, two parameters being the electric
charge $q_e$ and the magnetic charge $q_m$. Five asymptotic
parameters $M,\,D,\,\phi_\infty,\,Q_e,\,Q_m$ are functions of
$q_e,\,q_m$ which can be found numerically.

\subsection{Black holes on the linear dilaton background}
There is another type of physically interesting black hole solutions
which  asymptotically approach the linear dilaton background
(LDB)~\cite{Clement:2002mb}. For such black holes, some metric
functions diverge asymptotically, so to be able to recognize them
numerically we have to pass to some conformally rescaled
metric~\cite{Clement:2004ii}
\begin{equation}
ds^2_{\rm dual} = \mathrm{e}^{- 2 a (\phi - \phi_\infty)} \, ds^2,
\end{equation}
which has the following explicit form in our case
\begin{equation}
ds^2_{\rm dual} = P^{-1} \left( - w dt^2 + \frac{dr^2}{w} + \rho^2
d\Omega^2 \right).
\end{equation}

In the dual frame the asymptotic metric for LDB solutions could
be either $M_2 \times S^2$ (only for a special value of the dilation
coupling, $a^2 = 1$ for the 4-dimensional theory) or $AdS_2 \times
S^2$~\cite{Clement:2004ii, Clement:2005vn}. Therefore, if we rewrite
the metric in the following form
\begin{equation}
ds^2_{\rm dual} = - \frac{w}{P} dt^2 + \frac{P}{w} du^2 + P^{-1}
\rho^2 d\Omega^2,
\end{equation}
where $du = dr/P$, then the metric functions should have the
following asymptotic limit (determining the radii of the~$S^2$)
\begin{equation}\label{aLDBc}
P^{-1} \rho^2 \sim R_0, \qquad \frac{w}{P} \sim \tilde R_0^2 \; u^2,
\end{equation}
where $R_0$ and $\tilde R_0$ denote the curvature radii of the~$S^2$
and $AdS_2$ respectively. The differentiation of the second equation
gives the relation
\begin{equation}
\left( \frac{P}{w} \right)^{1/2} w' - \left( \frac{w}{P} \right)^{1/2} P'
\sim \tilde R_0,
\end{equation}
which can be used in the numerical procedure.

\section{Numerical analysis}
In this section we present numerical results for the global extremal
dyonic black hole solutions. We extend the local solutions
constructed via series expansions near the horizon to the asymptotic
region by numerical integration. As expected, the free parameter
$P_1$ turns out to be fixed by the asymptotic conditions, either of
asymptotic flatness or the linear dilaton asymptotic (LDB).

\subsection{Asymptotically flat dyons}
For the pure electric extremal black holes~\cite{Chen:2006ge}, the
global solutions were found to exits only in a limited range of the
dilaton couplings less than a critical value $a_{\rm cr}$. By
turning on the magnetic charge, the range of $a$ for domain of
existence can be extended from the interval $0 \le a < a_{\rm cr}$
to an infinite sequence of disconnected intervals, located between
the limiting curves $\Upsilon_i = 0$ in the two-dimensional
parameter plane $a, \, q_m$. More precisely, the regular solutions
exist for $a$ satisfying
\begin{equation}
a_i^-(q_m) < a < a_i^+(q_m),
\end{equation}
where $a_i^-$ is located above the curve $\Upsilon_{i-1} = 0$ and
$a_i^+$ below the curve $\Upsilon_i = 0$. The domains of existence
obtained numerically are shown in Fig.~1. The most surprising
feature of this plot is that for $q_m\to \infty$, which is
equivalent to turning off the Gauss-Bonnet term, the domain of
existence of dyon solution does not reduce to discrete values of the
dilaton coupling constant $a=a_i$. As we have seen, already the
series solutions in the near-horizon region are essentially
different in the EDGB model and in the curvature uncorrected EMD
theory: the former being one-parametric, while the latter ---
two-parametric. Transition between these two series solutions is
unclear, the limiting form of the series in the EDGB model depends
on the direction in the $a,\,q_m$ plane in which the limit $q_m\to
\infty$ is taken. Numerical solutions were checked to exist up to
some large values of $q_m$, that is, with small Gauss-Bonnet
coupling. They still exist for continuously varied $a$, not for
discrete values as in the pure EMD theory. Therefore the curvature
corrected theory gives qualitatively different predictions for dyons
even when the Gauss-Bonnet coupling is small.

Our numerical results reveal the following behavior of dyon
solutions for large  $q_m$ (or small Gauss-Bonnet coupling). If we
fix  $q_m$ and then take the dilaton coupling $a$ close to the
critical values $a_i^\pm$, the parameter $P_1$ (which  is fixed by
the asymptotic flatness conditions $w_\infty \to 1, \rho'_\infty \to
1$) is the going to diverge, and the solution becomes ill-defined.
However, if we fix the value $a\neq a_i$ of the dilaton coupling and
then move $q_m$ to infinity, the parameter $P_1$ approaches zero. In
this case, the limiting solution  $q_m \to \infty$ corresponds to
the dyonic {\em Reissner-Nordstr\"om } extremal black hole (with the
frozen dilaton). Such solutions, however, do not exist in the pure
EMD theory. Moreover, the subsequent limit $a^2 \to a_i^2$ always
gives the dyonic   Reissner-Nordstr\"om   solution.  Therefore, the
extremal dyons in the pure EMD theory look like a set of discrete
points in the two-dimensional continuum which is difficult to
resolve. Nevertheless, the limiting value of the radius of the
horizon is the same both in the EDGB and the EMD theories (see also
Appendix A).

Using the symmetry of the solution space under the
$\delta$-transformation (\ref{tqe}), one can generate the sequence
of solutions with different electric charges $q_e$ and
correspondingly with different masses, dilaton charges and the
asymptotic values of the dilaton $\phi_\infty$. Since variation of
the electric charge is essentially equivalent to variation of the
unique parameter $\rho_0$ (for fixed $q_m$) in the horizon
expansion, it is clear, that using $\delta$-transformation we will
generate {\em all} extremal solutions. Under this transformation the
mass and the dilaton charge scale as $\mathrm{e}^\delta$, while the
electric charge and the dilaton exponent $\mathrm{e}^{2a
\phi_\infty}$ scale as $\mathrm{e}^{2\delta}$. Therefore the ratios
\begin{equation}
k_M = \frac{M^2}{q_e}, \qquad k_D = \frac{D^2}{q_e}, \qquad k_\phi =
\frac{\mathrm{e}^{2a\phi_\infty} }{q_e},
\end{equation}
depend only on $a$. Their numerical plots are presented in Fig. 2.

With growing magnetic charge,  two end points of the allowed
interval of $a$ tend to   the boundary curves. When the dilaton
coupling approaches the critical values (the end points $a_i^\pm$ of
each segment of the regular solution with fixed $q_m$), the global
physical quantities such as the mass $M$, the dilaton charge $D$,
the electric charge $Q_e = q_e \, \mathrm{e}^{- a \, \phi_\infty}$
and the magnetic charge $Q_m = q_m \, \mathrm{e}^{a \, \phi_\infty}$
diverge and the only free parameter $P_1$ also goes to infinity (in
order to ensure an asymptotic flatness). However, the ratio of the
following physical quantities
\begin{equation}
k_{BPS} = \frac{1 + a^2}{2 a^2} \, \frac{a^2 M^2 + D^2}{Q_e^2 + Q_m^2},
%
\end{equation}
has a simple limit $k_{BPS} \to 1$ (Fig. 3). Remarkably, as  $a\to
a_i^\pm$, the associated global parameters tend to satisfy the
following relation
\begin{equation}\label{BPScond}
a^2 M^2 + D^2 = \frac{2 a^2}{ 1 + a^2 } \, (Q_e^2 + Q_m^2),
\end{equation}
which coincides with the BPS condition for charged black holes in
the EMD theory without curvature corrections. One can also see that
in this limit the discrete S-duality of the EMD theory is restored.
This feature is similar to that in another stringy generalization of
the EMD theory in which the Maxwell action is replaced by the
Born-Infeld action but no Gauss-Bonnet term is
introduced~\cite{Clement:2000ue}.

\begin{figure}[ht]
\includegraphics[width=10cm]{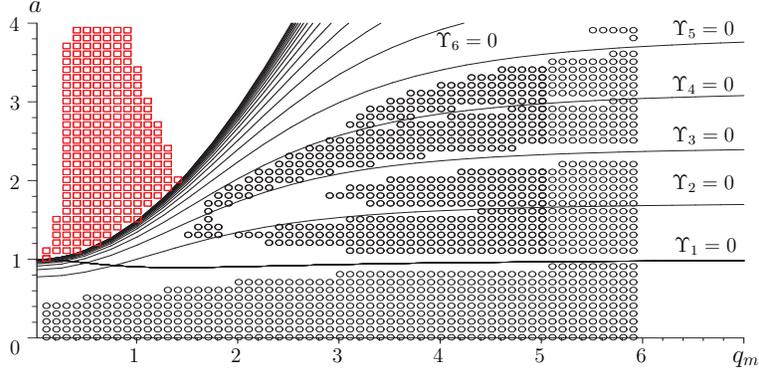}
\caption{The domains of existence of EDGB dyons in the two
dimensional parameter plane $a, \, q_m$ for $q_m\leq 6$. The family
of limiting curves corresponds to solutions of the equations
$\Upsilon_i = 0$ implying singularities of higher order coefficients
in the horizon series. Black circles correspond to asymptotically
flat solutions, red squares
--- to black holes on the linear dilaton background.}
\end{figure}

\begin{figure}[ht]
\includegraphics[width=14cm]{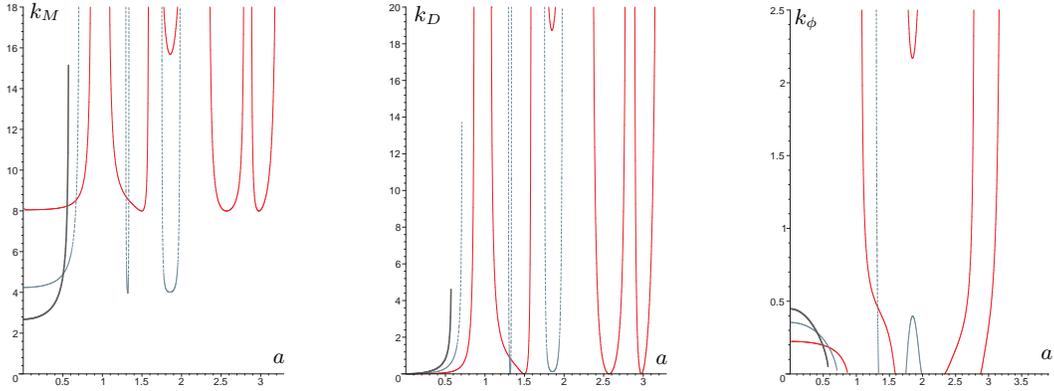}
\caption{Asymptotic parameters $M,\,D,\,\phi_\infty$ in terms of the
horizon parameters $q_e,\,q_m$ as given by the ratios $k_M =
\frac{M^2}{q_e}, k_D = \frac{D^2}{q_e}$ and $k_\phi =
\frac{\mathrm{e}^{2a\phi}}{q_e}$ depending on $q_m$. These ratios
are shown as functions of the dilaton coupling constant $a$ for some
values of the magnetic charge: $q_m = 1$ (black, thick), $q_m = 2$
(gray, dashed) and $q_m = 4$ (red).}
\end{figure}

\begin{figure}[ht]
\includegraphics[width=8cm]{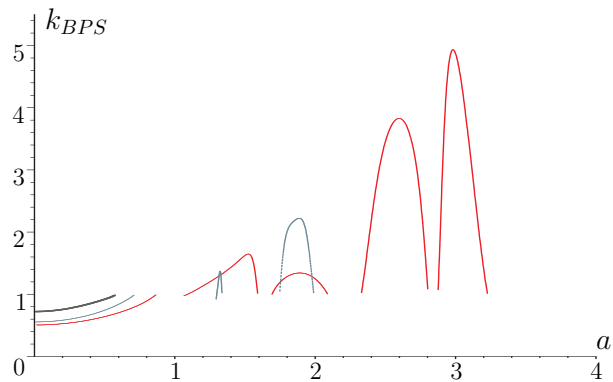}
\caption{The ratios $k_{BPS} = \frac{(1 + a^2) (a^2 M^2 + D^2)}{2
a^2 (Q_e^2 + Q_m^2)}$ as functions of $a$ for some values of
magnetic charge: $q_m = 1$ (black, thick), $q_m = 2$ (gray, dashed)
and $q_m = 4$ (red). One can see the approaching of the pseudo-BPS
limit at the end points of the existence domains.}
\end{figure}

\subsection{Magnetic black holes on the LDB background}
The domain of existence of black holes on the linear dilaton
background satisfying the condition (\ref{aLDBc}) is located to the
left from the condensing family of limiting curves $\Upsilon_i = 0$
on the parameter plane $a, \, q_m$ in Fig. 1. The boundary curve is
given by $\Upsilon_\infty = 0$ or, explicitly,
\begin{equation}
q_m^4 + 4 \alpha q_m^2 - 4 \alpha^2 (a^2 - 1) = 0.
\end{equation}
Domains of existence of all asymptotic flat solutions are located to
the right from this curve while those with the LDB asymptotics ---
to the left. Asymptotically LDB dyons do not exist for dilaton
couplings less than one. Moreover, for larger magnetic charges, the
asymptotically LDB solutions require larger value of the dilaton
coupling.

From an analysis of similar solutions in the EMD
theory~\cite{Clement:2005vn} it follows that one of the charges of
the dyonic configuration defines the strength of the background
electric or magnetic fields in the LDB, while another is associated
with the charge of a black hole. So one can have electric black
holes on the magnetic LDB or magnetic black holes on the electric
LDB. The asymptotic form of the metric is
\begin{equation}\label{mfi_ldb}
w \sim \frac{r}{\xi_0}, \qquad \rho \sim (\xi_0 r)^{\frac12},
\end{equation}
where $\xi_0$ is the scaling parameter associated with the field
strength of the background, and the dilaton  for the electric black
hole on a magnetic background is
\begin{equation}\label{dfi_ldb1}
P = \mathrm{e}^{2 a \phi} \sim \frac{r}{\xi_0},
\end{equation}
while for the magnetic black hole on an electric background
\begin{equation}\label{dfi_ldb2}
P = \mathrm{e}^{2 a \phi} \sim \frac{\xi_0}{r}.
\end{equation}
The solutions presented in this paper are consistent with the
dilaton field (\ref{dfi_ldb2}) and the scaling parameters $\mu = - 2
\delta = - \ln \xi_0$. Thus, our solutions can be interpreted as
magnetically charged black holes on the electric LDB (recall that
the discrete S-duality is broken in the EDGB theory). In the case of
the vanishing magnetic charge they are physically expected to reduce
to the pure electric LDB without a black hole. Technically, however,
our solutions can not have such a limit since we have assumed the
existence of the horizon a priori.

\section{Discussion}
In this paper, we have shown that the EDBG four-dimensional gravity
admits extremal dyonic black hole solutions with the horizon of the
$AdS_2 \times S^2$ type. Somewhat surprisingly, adding the
Gauss-Bonnet term to the Einstein-Maxwell-dilaton theory leads to an
enhancement of the domain of parameters for which the global
solutions exist. Namely, the asymptotically flat dyon solutions in
the EMD theory exist only for a discrete sequence of the dilaton
coupling constant values, while in the model with the Gauss-Bonnet
term the continuously varying parameters are allowed. An effective
parameter space is a two-dimensional plane which is split into the
sequence of regions separated by the limiting curves,  marking
singularities of the coefficients of the local power series
solutions. These curves are related to the above discrete sequence
of parameters of the EMD dyons, and when approaching them, the
solutions of the EDGB model exhibit  saturation features similar to
the BPS conditions of the EMD theory.

The relationship between the extremal EDGB dyons and those in the
EMD theory is non-trivial. Dyon solutions of the EDGB theory exist
only with non-zero electric charge, and for large values of the
magnetic charge they tend to the Reissner-Nordstr\"om solution with
a frozen dilaton and not to the discrete family of extremal dyons in
the EMD theory as could be expected. Therefore, the pure EMD theory
without curvature corrections predicts different black hole
solutions than the corresponding curvature-corrected theory in the
limit of vanishing curvature coupling. Remarkably, the latter
limiting theory leads the same value of the horizon radius, as the
pure EMD theory.

The entropy of the extremal EDGB dyons interpolates between the
Bekenstein-Hawking value in the limit of the large magnetic charges
(equivalent to the vanishing curvature coupling) and the doubled
Bekenstein-Hawking value in the limit of purely electric solutions.
The entropy can be calculated using only local analytical solutions
valid in the vicinity of the event horizon. The expression obtained,
however, does not bear any sign of the bounds on the parameters for
which the global black hole solutions exist. Such bounds can be
revealed once we try to extend the local solutions to infinity as
asymptotically flat ones. This lesson is worth to be kept in mind in
the  discussion of the entropy of the curvature corrected black
holes in string theory. Purely local analysis of classical solutions
is still insufficient to fully understand the entropy of black
holes.

We also found the second family of the EDGB dyons which
asymptotically approach the linear dilaton background. For them, the
metric is not asymptotically flat, but physical value of the black
hole mass and other parameters can be extracted by subtracting the
background values. In string theory such black holes correspond to
thermalized states of the quantum theory through the QFT/Domain wall
correspondence. Our solutions may be interpreted as magnetically
charged extremal black holes on the electric linear dilaton
background.

\section*{Acknowledgments}
DVG thanks the Department of Physics of NCU for hospitality and the
National Center of Theoretical Sciences and the Center for
Mathematics and Theoretical Physics at NCU for support during his
visit in January 2007. The work was also supported in part by the
RFBR grant 08-02-01398. CMC and DGO were supported by the National
Science Council of the R.O.C. under the grant NSC
96-2112-M-008-006-MY3. CMC was supported in part by the National
Center of Theoretical Sciences (NCTS) and the Center for Mathematics
and Theoretic Physics at NCU.

\begin{appendix}
\section{Extremal dyons in 4D Einstein-Maxwell-dilaton theory}
The action and the equations of motion of the four-dimensional
Einstein-Maxwell-dilaton (EMD) theory with arbitrary coupling $a$
are obtained setting $\alpha = 0$ in the
Eqs.~(\ref{Eqw}-\ref{Eqphi}). Consider again the series expansions
in the form (\ref{locdeg}) around the horizon $r = r_H$ in terms of
the deviation $x = r - r_H$. In the lowest order we obtain:
\begin{equation}
P_0 = \frac{q_e}{q_m}, \qquad \rho_0^2 = 2 q_e q_m,
\end{equation}
which coincides with the $\alpha \to 0$ limit of the corresponding
EDGB relations. However,  calculation of the higher order
coefficients shows that one deals with the local solution containing
{\em two} free parameters (with fixed electric and magnetic
charges), which can be conveniently chosen as $\rho_1$ and $P_1$.
Recall, that the corresponding local solution in the EDGB model
constructed in Sec.~3 contained only one free parameter $P_1$.
Another new feature consists in bifurcations encountered in the
calculation of the higher expansion coefficients as was described in
Sec.~3. These bifurcations give rise to a  sequence of discrete
values of the dilaton coupling constant $ a_i^2 = i (i+1)/2$ for
which only one obtains the non-trivial expansions. For each $i$ the
series expansion for the dilaton function starts with the term of
the $i$-th order, $P_i$, with all lower coefficients being zero,
$P_{j<i} = 0$ (except for $P_0$). This discrete sequence   was also
observed in the Ref.~\cite{Poletti:1995yq} where it was shown that
the asymptotically flat solutions exhibit {\sl two horizons} if $a =
0, 1, \sqrt3, \sqrt6, \cdots, \sqrt{n(n+1)/2}$, and one horizon
otherwise. Since the  solutions with the second order zero of the
metric function $g_{tt}$ at the horizon can be obtained only
starting with the non-extremal solutions with two horizons, this is
consistent with the special values we find here for genuinely
extremal solutions. Note that only in the first two cases $a = 1$
and $a = \sqrt3$  the exact dyonic solutions have been obtained
analytically. In what follows we show  how this can be done via
direct summation of the series solutions.

\subsection{$a^2 = 1$}
The series solution reads
\begin{eqnarray}
w(r) &=& \frac1{\rho_0^2} \, x^2 - \frac{2 \rho_1}{\rho_0^3} \, x^3 +
\left( \frac{3 \rho_1^2}{\rho_0^4} + \frac{P_1^2 q_m^4}{\rho_0^6} \right) x^4 -
\left( \frac{4 \rho_1^3}{\rho_0^5} + \frac{4 \rho_1 P_1^2 q_m^4}{\rho_0^7} \right)
x^5 + \left( \frac{5 \rho_1^4}{\rho_0^6} + \frac{10 \rho_1^2 P_1^2 q_m^4}{\rho_0^8}
+ \frac{P_1^4 q_m^8}{\rho_0^{10}} \right) x^6
\nonumber\\
&-& \left( \frac{6 \rho_1^5}{\rho_0^7} + \frac{20 \rho_1^3 P_1^2 q_m^4}{\rho_0^9}
+ \frac{6 \rho_1 P_1^4 q_m^8}{\rho_0^{11}} \right) x^7 + O(x^8),
\\
\rho(r) &=& \rho_0 + \rho_1 \, x - \frac{P_1^2 q_m^4}{2 \rho_0^3} \, x^2 +
\frac{\rho_1 P_1^2 q_m^4}{2 \rho_0^4} \, x^3 - \left( \frac{\rho_1^2 P_1^2 q_m^4}
{2 \rho_0^5} + \frac{P_1^4 q_m^8}{8 \rho_0^7} \right) x^4 +
\left( \frac{\rho_1^2 P_1^2 q_m^4}{2 \rho_0^6} + \frac{3 \rho_1 P_1^4 q_m^8}
{8  \rho_0^8} \right) x^5
\nonumber\\
&-& \left( \frac{\rho_1^4 P_1^2 q_m^4}{2 \rho_0^7} + \frac{6 \rho_1^2 P_1^4 q_m^8}
{8 \rho_0^9} + \frac{P_1^6 q_m^{12}}{16 \rho_0^{11}} \right) x^6 + O(x^7),
\\
P(r) &=& P_0 + P_1 \, x - P_1 \left( \frac{\rho_1}{\rho_0} - \frac{P_1 q_m^2}
{\rho_0^2} \right) x^2 + P_1 \left( \frac{\rho_1}{\rho_0} - \frac{P_1  q_m^2}
{\rho_0^2} \right)^2 x^3 - P_1 \left( \frac{\rho_1}{\rho_0} - \frac{P_1 q_m^2}
{\rho_0^2} \right)^3 x^4 + O(x^5).
\end{eqnarray}
The summation is performed by collecting terms of different order
in $\rho_1$ inside the series in terms of $P_1^2 q_m^4$. For
example, for $\rho(r)$ one can rewrite the series as
\begin{eqnarray}
\rho(r) &=& \rho_0 + \rho_1 \, x - \frac{P_1^2 q_m^4}{2 \rho_0^3} x^2 \left( 1 -
\frac{\rho_1}{\rho_0} x + \frac{\rho_1^2}{\rho_0^2} x^2 - \cdots \right) +
\frac{P_1^4 q_m^8}{8 \rho_0^7} x^4 \left( 1 - 3 \frac{\rho_1}{\rho_0} x +
6 \frac{\rho_1^2}{\rho_0^2} x^2 - \cdots \right) + \cdots
\nonumber\\
&=& (\rho_0 + \rho_1 \, x) \left( 1 - \frac{P_1^2 q_m^4}{2 \rho_0^2} \frac{x^2}
{(\rho_0 + \rho_1 x)^2} + \frac{P_1^4 q_m^8}{8 \rho_0^4} \frac{x^4}{(\rho_0 +
\rho_1 x)^4} - \cdots \right)
\nonumber\\
&=& \sqrt{(\rho_0 + \rho_1 \, x)^2 - \hat P_1^2 q_m^4 x^2}.
\end{eqnarray}
A similar pattern can be found in the series expansion for $w(r)$,
the result being as simple as
\begin{equation}
w(r) = \frac{x^2}{\rho^2(r)}.
\end{equation}
For $P(r)$, introducing a new parameter  $\bar P_1$ defined via $P_1 =
\bar P_1 \rho_0$ one gets just a geometric recurrence:
\begin{eqnarray}
P(r) &=& P_0 + P_1 x \left[ 1 - \left( \frac{\rho_1}{\rho_0} - \frac{P_1 q_m^2}
{\rho_0^2} \right) x + \left( \frac{\rho_1}{\rho_0} - \frac{P_1 q_m^2}{\rho_0^2}
\right)^2 x^2 - \left( \frac{\rho_1}{\rho_0} - \frac{P_1 q_m^2}{\rho_0^2}
\right)^3 x^3 + \cdots \right]
\\
&=& P_0 + \frac{\rho_0^2 \bar P_1 x}{(\rho_0 + \rho_1 x) - \bar P_1 q_m^2 x}
= P_0 \frac{(\rho_0 + \rho_1 x) + \bar P_1 q_m^2 x}{(\rho_0 + \rho_1 x) -
\bar P_1 q_m^2 x}.
\end{eqnarray}
Thus, we have obtained an  exact solution in a closed form.

\subsection{$a^2 = 3$}
This case corresponds to the Kaluza-Klein theory. The series
expansions are
\begin{eqnarray}
w(r) &=& \frac1{\rho_0^2} \, x^2 - \frac{2 \rho_1}{\rho_0^3} \, x^3 +
\frac{3 \rho_1^2}{\rho_0^4} \, x^4 - \frac{4 \rho_1^3}{\rho_0^5} \, x^5 +
\left( \frac{5 \rho_1^4}{\rho_0^6} + \frac{2 P_2^2 q_m^4}{9 \rho_0^6} \right) x^6
- \left( \frac{6 \rho_1^5}{\rho_0^7} + \frac{4 \rho_1 P_2^2 q_m^4}{3 \rho_0^7}
\right) x^7 + O(x^8),
\\
\rho(r) &=& \rho_0 + \rho_1 \, x - \frac{P_2^2 q_m^4}{9 \rho_0^3} \, x^4 +
\frac{\rho_1 P_2^2 q_m^4}{3 \rho_0^4} \, x^5 - \frac{2 \rho_1^2 P_2^2 q_m^4}
{3 \rho_0^5} \, x^6 + O(x^7),
\\
P(r) &=& P_0 + P_2 \, x^2 - \frac{2 \rho_1 P_2}{\rho_0} \, x^3 +
\frac{P_2 (P_2 q_m^2 + 3 \rho_1^2)}{\rho_0^2} \, x^4 + O(x^5) .
\end{eqnarray}
Rearranging an expansion for $\rho(r)$ in the same way we obtain
the exact sum
\begin{equation}
\rho(r) = \left[ (\rho_0 + \rho_1 x)^4 - \frac49 P_2^2 q_m^4 x^4 \right]^{1/4}.
\end{equation}
The function $w$ is given again by  $w(r) = x^2/\rho^2(r)$. However,
it is more difficult to sum up the series expansion for the dilaton
function  $P(r)$. An easier way to find $P(r)$ is to solve the
equations of motion directly using the above results for $\rho(r)$
and $w(r)$. This gives
\begin{equation}
P(r) = P_0 \left[ \frac{(\rho_0 + \rho_1 x)^2 + \frac23 P_2 q_m^2 x^2}{(\rho_0
+ \rho_1 x)^2 - \frac23 P_2 q_m^2 x^2} \right]^{3/2}.
\end{equation}
Again, we have obtained an exact solution from the near horizon
expansion.

\subsection{$a^2 = 6$ and beyond}
One can try to repeat the same procedure for  higher $a_i$. In
particular, for $a^2 = 6\; (i = 3)$ one has the following series
expansions:
\begin{eqnarray}
w(r) &=& \frac1{\rho_0^2} \, x^2 - \frac{2 \rho_1}{\rho_0^3} \, x^3 +
\frac{3 \rho_1^2}{\rho_0^4} \, x^4 - \frac{4 \rho_1^3}{\rho_0^5} \, x^5
+ \frac{5 \rho_1^4}{\rho_0^6} \, x^6 - \frac{6 \rho_1^5}{\rho_0^7} \, x^7
\nonumber\\
&& + \left( \frac{7 \rho_1^6}{\rho_0^8} + \frac{P_3^2 q_m^4}{10 \rho_0^6} \right)
x^8 - \left( \frac{8 \rho_1^7}{\rho_0^9} + \frac{4 \rho_1 P_3^2 q_m^4 }
{5 \rho_0^7} \right) x^9 + O(x^{10}),
\nonumber\\
\rho(r) &=& \rho_0 + \rho_1 \, x - \frac{P_3^2 q_m^4}{20 \rho_0^3} \, x^6 +
\frac{\rho_1 P_3^2 q_m^4}{4 \rho_0^4} \, x^7 + O(x^8),
\nonumber\\
P(r) &=& P_0 + P_3 \, x^3 - \frac{3 P_3 \rho_1}{\rho_0} \, x^4 +
\frac{6 P_3 \rho_1^2}{\rho_0^2} \, x^5 + O(x^6).
\end{eqnarray}
Note, that the function $w(r)$ differs from  the
Reissner-Nordstr\"om $w(r)$ only in terms of the order $x^8$ and
higher (for higher $i$ from the terms of the order $2i+2$). For the
dilaton function an expansion starts from the cubic term (generally,
from the $i$-th terms). But we were not able to find a closed form
for these expansions as a whole.

\section{Higher-order terms in the local solution near the horizon}
Here we list some higher order coefficients in the near horizon
expansions for extremal black holes in the EDGB theory considered in
Sec.~3. The parameters  are $\rho_0$ which is determined by the
electric and magnetic charges, and $P_1$ (free) which must be fixed
by asymptotic conditions:
\begin{eqnarray}
P_0 &=& \frac{\rho_0^2}{2 (2 \alpha + q_m^2)},
\\
P_2 &=& \frac{P_1^2}{4 \alpha \rho_0^2 a^2 [(a^2 - 3) q_m^4 + 6 \alpha (a^2 - 2)
q_m^2 + 4 \alpha^2 (5 a^2 - 3)]} \Bigl[- (a^2 - 3) (a^2 - 1) q_m^8
\nonumber\\
&& + 4 \alpha (5 a^2 - 6) q_m^6 + 4 \alpha^2 (5 a^4 + 3 a^2 - 18) q_m^4 -
32 \alpha^3 (2 a^2 + 3) q_m^2 + 16 \alpha^4 (a^4 - 5 a^2 - 3) \Bigr],
\\
P_3 &=& \frac{P_1^3}{48 \alpha^2 a^4 \rho_0^4 [(a^2 - 3) q_m^4 + 6 \alpha (a^2 - 2)
q_m^2 + 4 \alpha^2 (5 a^2 - 3)] [(a^2 - 6) q_m^4 + 6 \alpha (a^2 - 4) q_m^2 +
4 \alpha^2 (8 a^2 - 6)]}
\nonumber\\
&& \Bigl[ - 3 (a^2 - 1)^2 (a^2 - 3) (a^2 + 4) q_m^{16} - 6 \alpha (a^2 - 1)
(24 a^6 - 91 a^4 - 9 a^2 + 96) q_m^{14}
\nonumber\\
&& - 4 \alpha^2 (317 a^8 - 1828 a^6 + 2444 a^4 + 45 a^2 - 1008) q_m^{12} -
8 \alpha^3 (483 a^8 - 3006 a^6 + 6283 a^4 - 2145 a^2 - 2016) q_m^{10}
\nonumber\\
&& - 16 \alpha^4 (355 a^8 - 960 a^6 + 6647 a^4 - 5625 a^2 - 2520) q_m^8 +
32 \alpha^5 (165 a^8 - 1762 a^6 - 1844 a^4 + 6975 a^2 + 2016) q_m^6 \nonumber\\
&& + 64 \alpha^6 (1156 a^8 - 1361 a^6 + 1922 a^4 + 4785 a^2 + 1008) q_m^4 +
128 \alpha^7 (369 a^8 - 49 a^6 + 1573 a^4 + 1755 a^2 + 288) q_m^2
\nonumber\\
&& + 512 \alpha^8 (23 a^8 + 55 a^6 + 165 a^4 + 135 a^2 + 18) \Bigr];
\end{eqnarray}
\begin{eqnarray}
\rho_1 &=& \frac{P_1}{4 \alpha a^2 \rho_0} \Bigl[ (a^2 - 1) q_m^4 +
2 \alpha (3 a^2 - 2) q_m^2 + 4 \alpha^2  (a^2 - 1) \Bigr],
\\
\rho_2 &=& \frac{P_1^2}{2 a^2 \rho_0^3 [(a^2 - 3) q_m^4 + 6 \alpha (a^2 - 2)
q_m^2 + 4 \alpha^2 (5 a^2 - 3)]} \Bigl[ (a^2 - 3) (a^2 - 2) q_m^8
\nonumber\\
&& - 6 \alpha (5 a^2 - 8) q_m^6 - 4 \alpha^2 (5 a^4 + 15 a^2 - 36) q_m^4 -
8 \alpha^3 (5 a^2 - 24) q_m^2 - 16 \alpha^4 (a^4 - 6) \Bigr],
\\
\rho_3 &=& \frac{P_1^3}{24 \alpha a^4 \rho_0^5 [(a^2 - 3) q_m^4 + 6
\alpha (a^2 - 2) q_m^2 + 4 \alpha^2 (5 a^2 - 3)] [(a^2 - 6) q_m^4 +
6 \alpha (a^2 - 4) q_m^2 + 4 \alpha^2 (8 a^2 - 6)]}
\nonumber\\
&&
\Bigl[ (a^2 - 1) (a^2 - 3) (2 a^4 + 19 a^2 - 36) q_m^{16} + 2 \alpha
(66 a^8 - 259 a^6 - 206 a^4 + 1293 a^2 - 864) q_m^{14}
\nonumber\\
&& + 4 \alpha^2 (304 a^8 - 1613 a^6 + 1085 a^4 + 3327 a^2 - 3024) q_m^{12} +
8 \alpha^3 (486 a^8 - 2716 a^6 + 3599 a^4 + 4065 a^2 - 6048) q_m^{10} \nonumber\\
&& + 16 \alpha^4 (438 a^8 - 687 a^6 + 2704 a^4 + 1605 a^2 - 7560) q_m^8 -
32 \alpha^5 (24 a^8 - 2273 a^6 + 2200 a^4 + 1689 a^2 + 6048) q_m^6
\nonumber\\
&& - 64 \alpha^6 (1098 a^8 - 2261 a^6 + 4467 a^4 + 2427 a^2 + 3024) q_m^4 -
128 \alpha^7 (342 a^8 - 758 a^6 + 2409 a^4 + 1173 a^2 + 864) q_m^2
\nonumber\\
&& - \alpha^8 (9728 a^8 - 21504 a^6 + 110592 a^4 + 53760 a^2 +
27648) \Bigr];
\end{eqnarray}
\begin{eqnarray}
w_2 &=& \frac1{\rho_0^2},
\\
w_3 &=& - \frac{P_1}{6 \alpha a^2 \rho_0^4} \Bigl[ 3 (a^2 - 1) q_m^4 + 6 \alpha
(3 a^2 - 2) q_m^2 + 4 \alpha^2 (5 a^2 - 3) \Bigr],
\\
w_4 &=& \frac{P_1^2}{48 \alpha^2 a^4 \rho_0^6 [(a^2 - 3) q_m^4 + 6 \alpha (a^2 - 2)
 q_m^2 + 4 \alpha^2 (5 a^2 - 3) ]} \Bigl[ 9 (a^2 - 3) (a^2 - 1)^2 q_m^{12}
\nonumber\\
&& + 18 \alpha (a^2 - 1) (9 a^4 - 31 a^2 + 18) q_m^{10} + 12 \alpha^2 (106 a^6 +
411 a^2 - 396 a^4 - 135) q_m^8
\nonumber\\
&& + 8 \alpha^3 (741 a^6 - 2150 a^4 + 1782 a^2 - 540) q_m^6 + 16 \alpha^4
(1010 a^6 - 2220 a^4 + 1413 a^2 - 405) q_m^4
\nonumber\\
&& + 96 \alpha^5 (223 a^6 - 404 a^4 + 195 a^2 - 54) q^2 + 64
\alpha^6 (173 a^6 - 269 a^4 + 99 a^2 - 27) \Bigr].
\end{eqnarray}
An inspection of these and higher-order coefficients shows the
systematic appearance of the combinations $\Upsilon_i$ in
denominators, as described in the Sec.~3.

\end{appendix}



\begin{references}

\bibitem{Chen:2006ge}
  C.~M.~Chen, D.~V.~Gal'tsov and D.~G.~Orlov,
  ``Extremal black holes in D = 4 Gauss-Bonnet gravity,''
  Phys.\ Rev.\  D {\bf 75}, 084030 (2007)
  [arXiv:hep-th/0701004].


\bibitem{deWit:2005ya}
  B.~de Wit,
  ``Supersymmetric black holes,''
  Fortsch.\ Phys.\  {\bf 54}, 183 (2006)
  [arXiv:hep-th/0511261].

\bibitem{Mohaupt:2005jd}
  T.~Mohaupt,
  ``Strings, higher curvature corrections, and black holes,''
  arXiv:hep-th/0512048.


\bibitem{Wald:1993nt}
  R.~M.~Wald,
  ``Black hole entropy in the Noether charge,''
  Phys.\ Rev.\ D {\bf 48}, 3427 (1993)
  [arXiv:gr-qc/9307038].

\bibitem{Jacobson:1993vj}
  T.~Jacobson, G.~Kang and R.~C.~Myers,
  ``On black hole entropy,''
  Phys.\ Rev.\ D {\bf 49}, 6587 (1994)
  [arXiv:gr-qc/9312023].

\bibitem{Iyer:1994ys}
  V.~Iyer and R.~M.~Wald,
   ``Some properties of Noether charge and a proposal for dynamical
   black hole entropy,''
  Phys.\ Rev.\ D {\bf 50}, 846 (1994)
  [arXiv:gr-qc/9403028].

\bibitem{Jacobson:1994qe}
  T.~Jacobson, G.~Kang and R.~C.~Myers,
  ``Black hole entropy in higher curvature gravity,''
  arXiv:gr-qc/9502009.


\bibitem{Myers:1998gt}
  R.~C.~Myers,
  ``Black holes in higher curvature gravity,''
  arXiv:gr-qc/9811042.

\bibitem{Callan:1988hs}
  C.~G.~Callan, R.~C.~Myers and M.~J.~Perry,
  ``Black holes in string theory,''
  Nucl.\ Phys.\ B {\bf 311}, 673 (1989).


\bibitem{Behrndt:1998eq}
  K.~Behrndt, G.~Lopes Cardoso, B.~de Wit, D.~Lust, T.~Mohaupt
  and W.~A.~Sabra,
   ``Higher-order black-hole solutions in N = 2 supergravity and
   Calabi-Yau string backgrounds,''
  Phys.\ Lett.\ B {\bf 429}, 289 (1998)
  [arXiv:hep-th/9801081].

\bibitem{LopesCardoso:1998wt}
  G.~Lopes Cardoso, B.~de Wit and T.~Mohaupt,
  ``Corrections to macroscopic supersymmetric black-hole entropy,''
  Phys.\ Lett.\ B {\bf 451}, 309 (1999)
  [arXiv:hep-th/9812082].

\bibitem{LopesCardoso:1999cv}
  G.~Lopes Cardoso, B.~de Wit and T.~Mohaupt,
  ``Deviations from the area law for supersymmetric black holes,''
  Fortsch.\ Phys.\  {\bf 48}, 49 (2000)
  [arXiv:hep-th/9904005].

\bibitem{LopesCardoso:1999ur}
  G.~Lopes Cardoso, B.~de Wit and T.~Mohaupt,
   ``Macroscopic entropy formulae and non-holomorphic corrections 
   for supersymmetric black holes,''
  Nucl.\ Phys.\ B {\bf 567}, 87 (2000)
  [arXiv:hep-th/9906094].

\bibitem{LopesCardoso:1999xn}
  G.~Lopes Cardoso, B.~de Wit and T.~Mohaupt,
  ``Area law corrections from state counting and supergravity,''
  Class.\ Quant.\ Grav.\  {\bf 17}, 1007 (2000)
  [arXiv:hep-th/9910179].

\bibitem{Mohaupt:2000mj}
  T.~Mohaupt,
  ``Black hole entropy, special geometry and strings,''
  Fortsch.\ Phys.\  {\bf 49}, 3 (2001)
  [arXiv:hep-th/0007195].

\bibitem{LopesCardoso:2000qm}
  G.~Lopes Cardoso, B.~de Wit, J.~Kappeli and T.~Mohaupt,
  ``Stationary BPS solutions in N = 2 supergravity with R**2 interactions,''
  JHEP {\bf 0012}, 019 (2000)
  [arXiv:hep-th/0009234].

\bibitem{LopesCardoso:2000fp}
  G.~Lopes Cardoso, B.~de Wit, J.~Kappeli and T.~Mohaupt,
  ``Examples of stationary BPS solutions in N = 2 supergravity theories 
  with R**2-interactions,''
  Fortsch.\ Phys.\  {\bf 49}, 557 (2001)
  [arXiv:hep-th/0012232].

\bibitem{Dabholkar:2004yr}
  A.~Dabholkar,
  ``Exact counting of black hole microstates,''
  Phys.\ Rev.\ Lett.\  {\bf 94}, 241301 (2005)
  [arXiv:hep-th/0409148].

\bibitem{Dabholkar:2004dq}
  A.~Dabholkar, R.~Kallosh and A.~Maloney,
  ``A stringy cloak for a classical singularity,''
  JHEP {\bf 0412}, 059 (2004)
  [arXiv:hep-th/0410076].

\bibitem{Sen:2004dp}
  A.~Sen,
  ``How does a fundamental string stretch its horizon?,''
  JHEP {\bf 0505}, 059 (2005)
  [arXiv:hep-th/0411255].

\bibitem{Hubeny:2004ji}
  V.~Hubeny, A.~Maloney and M.~Rangamani,
  ``String-corrected black holes,''
  JHEP {\bf 0505}, 035 (2005)
  [arXiv:hep-th/0411272].

\bibitem{Bak:2005mt}
  D.~Bak, S.~Kim and S.~J.~Rey,
  ``Exactly soluble BPS black holes in higher curvature N = 2 supergravity,''
  arXiv:hep-th/0501014.


\bibitem{Goldstein:2005hq}
  K.~Goldstein, N.~Iizuka, R.~P.~Jena and S.~P.~Trivedi,
  ``Non-supersymmetric attractors,''
  Phys.\ Rev.\ D {\bf 72}, 124021 (2005)
  [arXiv:hep-th/0507096].

\bibitem{Kallosh:2005ax}
  R.~Kallosh,
  ``New attractors,''
  JHEP {\bf 0512}, 022 (2005)
  [arXiv:hep-th/0510024].

\bibitem{Tripathy:2005qp}
  P.~K.~Tripathy and S.~P.~Trivedi,
  ``Non-supersymmetric attractors in string theory,''
  JHEP {\bf 0603}, 022 (2006)
  [arXiv:hep-th/0511117].

\bibitem{Giryavets:2005nf}
  A.~Giryavets,
  ``New attractors and area codes,''
  JHEP {\bf 0603}, 020 (2006)
  [arXiv:hep-th/0511215].

\bibitem{Goldstein:2005rr}
  K.~Goldstein, R.~P.~Jena, G.~Mandal and S.~P.~Trivedi,
  ``A C-function for non-supersymmetric attractors,''
  JHEP {\bf 0602}, 053 (2006)
  [arXiv:hep-th/0512138].

\bibitem{Kallosh:2006bt}
  R.~Kallosh, N.~Sivanandam and M.~Soroush,
  ``The non-BPS black hole attractor equation,''
  JHEP {\bf 0603}, 060 (2006)
  [arXiv:hep-th/0602005].

\bibitem{Kallosh:2006bx}
  R.~Kallosh,
  ``From BPS to non-BPS black holes canonically,''
  arXiv:hep-th/0603003.

\bibitem{Prester:2005qs}
  P.~Prester,
  ``Lovelock type gravity and small black holes in heterotic string theory,''
  JHEP {\bf 0602}, 039 (2006)
  [arXiv:hep-th/0511306].

\bibitem{Cvitan:2007pk}
  M.~Cvitan, P.~D.~Prester, A.~Ficnar, S.~Pallua and I.~Smolic,
  ``Five-dimensional black holes in heterotic string theory,''
  Fortsch.\ Phys.\  {\bf 56}, 406 (2008)
  [arXiv:0711.4962 [hep-th]].

\bibitem{Cvitan:2007hu}
  M.~Cvitan, P.~D.~Prester and A.~Ficnar,
  ``$\alpha'^2$-corrections to extremal dyonic black holes in heterotic string
  theory,''
  JHEP {\bf 0805}, 063 (2008)
  [arXiv:0710.3886 [hep-th]].

\bibitem{Cvitan:2007en}
  M.~Cvitan, P.~D.~Prester, S.~Pallua and I.~Smolic,
  ``Extremal black holes in D=5: SUSY vs. Gauss-Bonnet corrections,''
  JHEP {\bf 0711}, 043 (2007)
  [arXiv:0706.1167 [hep-th]].

\bibitem{Alishahiha:2006ke}
  M.~Alishahiha and H.~Ebrahim,
  ``Non-supersymmetric attractors and entropy function,''
  JHEP {\bf 0603}, 003 (2006)
  [arXiv:hep-th/0601016].

\bibitem{Sinha:2006yy}
  A.~Sinha and N.~V.~Suryanarayana,
  ``Extremal single-charge small black holes: Entropy function analysis,''
  Class.\ Quant.\ Grav.\  {\bf 23}, 3305 (2006)
  [arXiv:hep-th/0601183].

\bibitem{Chandrasekhar:2006kx}
  B.~Chandrasekhar, S.~Parvizi, A.~Tavanfar and H.~Yavartanoo,
  ``Non-supersymmetric attractors in R**2 gravities,''
  JHEP {\bf 0608}, 004 (2006)
  [arXiv:hep-th/0602022].

\bibitem{Parvizi:2006uz}
  S.~Parvizi and A.~Tavanfar,
   ``Partition function of non-supersymmetric black holes in the
   supergravity limit,''
  arXiv:hep-th/0602292.

\bibitem{Sahoo:2006rp}
  B.~Sahoo and A.~Sen,
   ``Higher derivative corrections to non-supersymmetric extremal
   black holes in N = 2 supergravity,''
  JHEP {\bf 0609}, 029 (2006)
  [arXiv:hep-th/0603149].

\bibitem{Astefanesei:2006sy}
  D.~Astefanesei, K.~Goldstein and S.~Mahapatra,
   ``Moduli and (un)attractor black hole thermodynamics,''
  arXiv:hep-th/0611140.


\bibitem{Sen:2005wa}
  A.~Sen,
   ``Black hole entropy function and the attractor mechanism in higher
   derivative gravity,''
  JHEP {\bf 0509}, 038 (2005)
  [arXiv:hep-th/0506177].

\bibitem{Sen:2005iz}
  A.~Sen,
  ``Entropy function for heterotic black holes,''
  JHEP {\bf 0603}, 008 (2006)
  [arXiv:hep-th/0508042].

\bibitem{Sen:2007qy}
  A.~Sen,
  ``Black Hole Entropy Function, Attractors and Precision Counting of
  Microstates,''
  arXiv:0708.1270 [hep-th].



\bibitem{Gibbons:1982ih}
  G.~W.~Gibbons,
  ``Antigravitating black hole solitons with scalar hair in N=4 supergravity,''
  Nucl.\ Phys.\ B {\bf 207} (1982) 337.

\bibitem{Gibbons:1987ps}
  G.~W.~Gibbons and K.~I.~Maeda,
  ``Black holes and membranes in higher dimensional theories with dilaton fields,''
  Nucl.\ Phys.\ B {\bf 298}, 741 (1988).

\bibitem{Gibbons:1985ac}
  G.~W.~Gibbons and D.~L.~Wiltshire,
  ``Black holes In Kaluza-Klein theory,''
  Annals Phys.\  {\bf 167}, 201 (1986)
  [Erratum-ibid.\  {\bf 176}, 393 (1987)].

\bibitem{Garfinkle:1990qj}
  D.~Garfinkle, G.~T.~Horowitz and A.~Strominger,
  ``Charged black holes in string theory,''
  Phys.\ Rev.\ D {\bf 43}, 3140 (1991)
  [Erratum-ibid.\ D {\bf 45}, 3888 (1992)].

\bibitem{Cornalba:2006hc}
  L.~Cornalba, M.~S.~Costa, J.~Penedones and P.~Vieira,
  ``From fundamental strings to small black holes,''
  JHEP {\bf 0612} (2006) 023
  [arXiv:hep-th/0607083].
  
\bibitem{Mignemi:1992nt}
  S.~Mignemi and N.~R.~Stewart,
  ``Charged black holes in effective string theory,''
  Phys.\ Rev.\ D {\bf 47}, 5259 (1993)
  [arXiv:hep-th/9212146].

\bibitem{Mignemi:1993ce}
  S.~Mignemi,
  ``Dyonic black holes in effective string theory,''
  Phys.\ Rev.\ D {\bf 51}, 934 (1995)
  [arXiv:hep-th/9303102].


\bibitem{Kanti:1995vq}
  P.~Kanti, N.~E.~Mavromatos, J.~Rizos, K.~Tamvakis and E.~Winstanley,
  ``Dilatonic Black Holes in Higher Curvature String Gravity,''
  Phys.\ Rev.\ D {\bf 54}, 5049 (1996)
  [arXiv:hep-th/9511071].

\bibitem{Torii:1996yi}
  T.~Torii, H.~Yajima and K.~I.~Maeda,
  ``Dilatonic black holes with Gauss-Bonnet term,''
  Phys.\ Rev.\ D {\bf 55}, 739 (1997)
  [arXiv:gr-qc/9606034].

\bibitem{Alexeev:1996vs}
  S.~O.~Alexeev and M.~V.~Pomazanov,
  ``Black hole solutions with dilatonic hair in higher curvature gravity,''
  Phys.\ Rev.\ D {\bf 55}, 2110 (1997)
  [arXiv:hep-th/9605106].

\bibitem{Alexeev:1997ua}
  S.~O.~Alexeev and M.~V.~Pomazanov,
  ``Singular regions in black hole solutions in higher order curvature gravity,''
  arXiv:gr-qc/9706066.

\bibitem{Guo:2008hf}
  Z.~K.~Guo, N.~Ohta and T.~Torii,
  ``Black Holes in the Dilatonic Einstein-Gauss-Bonnet Theory in Various
  Dimensions I -- Asymptotically Flat Black Holes --,''
  arXiv:0806.2481 [gr-qc].

\bibitem{Melis:2005xt}
  M.~Melis and S.~Mignemi,
  ``Global properties of dilatonic Gauss-Bonnet black holes,''
  Class.\ Quant.\ Grav.\  {\bf 22}, 3169 (2005)
  [arXiv:gr-qc/0501087].

\bibitem{Melis:2005ji}
  M.~Melis and S.~Mignemi,
  ``Global properties of charged dilatonic Gauss-Bonnet black holes,''
  Phys.\ Rev.\ D {\bf 73}, 083010 (2006)
  [arXiv:gr-qc/0512132].

\bibitem{Mignemi:2006ut}
  S.~Mignemi,
  ``Black hole solutions of dimensionally reduced Einstein-Gauss-Bonnet
  gravity,''
  Phys.\ Rev.\  D {\bf 74}, 124008 (2006)
  [arXiv:gr-qc/0607005].

\bibitem{Melis:2006fj}
  M.~Melis and S.~Mignemi,
  ``Black hole solutions of dimensionally reduced Einstein-Gauss-Bonnet
  gravity with a cosmological constant,''
  Phys.\ Rev.\  D {\bf 75}, 024042 (2007)
  [arXiv:gr-qc/0609133].


\bibitem{Torii:1998gm}
  T.~Torii and K.~I.~Maeda,
  ``Stability of a dilatonic black hole with a Gauss-Bonnet term,''
  Phys.\ Rev.\ D {\bf 58}, 084004 (1998).

\bibitem{Dotti:2004sh}
  G.~Dotti and R.~J.~Gleiser,
   ``Gravitational instability of Einstein-Gauss-Bonnet black holes under
   tensor mode perturbations,''
  Class.\ Quant.\ Grav.\  {\bf 22}, L1 (2005)
  [arXiv:gr-qc/0409005].

\bibitem{Dotti:2005sq}
  G.~Dotti and R.~J.~Gleiser,
   ``Linear stability of Einstein-Gauss-Bonnet static spacetimes. I:
   Tensor perturbations,''
  Phys.\ Rev.\ D {\bf 72}, 044018 (2005)
  [arXiv:gr-qc/0503117].

\bibitem{Gleiser:2005ra}
  R.~J.~Gleiser and G.~Dotti,
   ``Linear stability of Einstein-Gauss-Bonnet static spacetimes. II:
   Vector and scalar perturbations,''
  Phys.\ Rev.\ D {\bf 72}, 124002 (2005)
  [arXiv:gr-qc/0510069].

\bibitem{Moura:2006pz}
  F.~Moura and R.~Schiappa,
  ``Higher-derivative corrected black holes: Perturbative stability and
  absorption cross-section in heterotic string theory,''
  Class.\ Quant.\ Grav.\  {\bf 24}, 361 (2007)
  [arXiv:hep-th/0605001].



\bibitem{Poletti:1995yq}
  S.~J.~Poletti, J.~Twamley and D.~L.~Wiltshire,
  ``Dyonic dilaton black holes,''
  Class.\ Quant.\ Grav.\  {\bf 12}, 1753 (1995)
  [Erratum-ibid.\  {\bf 12}, 2355 (1995)]
  [arXiv:hep-th/9502054].


\bibitem{Clement:2002mb}
  G.~Clement, D.~Gal'tsov and C.~Leygnac,
  ``Linear dilaton black holes,''
  Phys.\ Rev.\  D {\bf 67}, 024012 (2003)
  [arXiv:hep-th/0208225].


\bibitem{Donets:1995ya}
  E.~E.~Donets and D.~V.~Gal'tsov,
  ``Stringy sphalerons and Gauss-Bonnet term,''
  Phys.\ Lett.\ B {\bf 352}, 261 (1995)
  [arXiv:hep-th/9503092].


\bibitem{Chen:2008px}
  C.~M.~Chen,
  ``Extremal dilatonic black holes in 4D Gauss-Bonnet gravity,''
  Prog.\ Theor.\ Phys.\ Suppl.\  {\bf 172}, 161 (2008)
  [arXiv:0801.0032 [hep-th]].


\bibitem{Cai:2007cz}
  R.~G.~Cai, C.~M.~Chen, K.~i.~Maeda, N.~Ohta and D.~W.~Pang,
  ``Entropy function and universality of entropy-area relation for
  small black holes,''
  Phys.\ Rev.\  D {\bf 77}, 064030 (2008)
  [arXiv:0712.4212 [hep-th]].


\bibitem{Clement:2004ii}
  G.~Clement, D.~Gal'tsov and C.~Leygnac,
  ``Black branes on the linear dilaton background,''
  Phys.\ Rev.\  D {\bf 71}, 084014 (2005)
  [arXiv:hep-th/0412321].

\bibitem{Clement:2005vn}
  G.~Clement, D.~Gal'tsov, C.~Leygnac and D.~Orlov,
  ``Dyonic branes and linear dilaton background,''
  Phys.\ Rev.\  D {\bf 73}, 045018 (2006)
  [arXiv:hep-th/0512013].


\bibitem{Clement:2000ue}
  G.~Clement and D.~Gal'tsov,
   ``Solitons and black holes in Einstein-Born-Infeld-dilaton theory,''
  Phys.\ Rev.\ D {\bf 62}, 124013 (2000)
  [arXiv:hep-th/0007228].

\end{references}
\end{document}